\documentclass[10pt,letterpaper]{article}
\usepackage[top=0.85in,left=2.25in,footskip=0.75in]{geometry}
\usepackage{graphicx} % Required for inserting images
\usepackage{amsmath} % assumes amsmath package installed
\usepackage{amssymb}
\usepackage{amsthm,setspace}
\usepackage[utf8x]{inputenc}

\theoremstyle{plain}

\usepackage{dblfloatfix}
\usepackage{dsfont}
\usepackage{setspace}
\usepackage{hyperref}
\usepackage{xcolor}
\usepackage{cite}
\usepackage{soul}
\usepackage{bbm}
\usepackage[most]{tcolorbox}

\usepackage{changepage}

\usepackage[utf8x]{inputenc}

\usepackage{textcomp,marvosym}

\usepackage{cite}
\usepackage{soul}

\usepackage{nameref,hyperref}

\usepackage[right]{lineno}

\usepackage{setspace}
\usepackage{microtype}
\DisableLigatures[f]{encoding = *, family = * }

\usepackage{array}
\usepackage{todonotes}

\newcolumntype{+}{!{\vrule width 2pt}}

\newlength\savedwidth

\raggedright
\usepackage[aboveskip=1pt,labelfont=bf,labelsep=period,justification=raggedright,singlelinecheck=off]{caption}

\makeatletter
\renewcommand{\@biblabel}[1]{\quad#1.}
\makeatother

\usepackage{lastpage,fancyhdr,graphicx}
\usepackage{epstopdf}
\fancyheadoffset[L]{2.25in}
\fancyfootoffset[L]{2.25in}

\begin{document}
\vspace*{0.2in}

\begin{flushleft}
{\Large
\textbf\newline{Alertness Optimization for Shift Workers Using \\ a Physiology-based Mathematical Model}
}
\newline
\\
Zidi Tao\textsuperscript{1,2*},
A. Agung Julius\textsuperscript{1,2},
John T. Wen\textsuperscript{1,2},
\\
\bigskip
\textbf{1} Dept. Electrical, Computer, and Systems Engineering, Rensselaer Polytechnic Institute, Troy, NY, USA
\\
\textbf{2} Lighting Enabled Systems and Applications (LESA) Engineering Research Center, Rensselaer Polytechnic Institute, Troy, NY, USA
\\
%\textbf{3} Dept. Industrial and Systems Engineering, Rensselaer Polytechnic Institute, Troy, NY, USA
%\\
\bigskip

% Use the asterisk to denote corresponding authorship and provide email address in note below.
* Corresponding Author: taoz2@rpi.edu

\end{flushleft}

\section*{Abstract}

Sleep is vital for maintaining cognitive function, facilitating metabolic waste removal, and supporting memory consolidation. However, modern societal demands, particularly shift work, often disrupt natural sleep patterns. This can induce excessive sleepiness among shift workers in critical sectors such as healthcare and transportation and increase the risk of accidents. The primary contributors to this issue are misalignments of circadian rhythms and enforced sleep-wake schedules. 

Regulating circadian rhythms that are tied to alertness can be regarded as a control problem with control inputs in the form of light and sleep schedules. In this paper, we address the problem of optimizing alertness by optimizing light and sleep schedules to improve the cognitive performance of shift workers. A key tool in our approach is a mathematical model that relates the control input variables (sleep and lighting schedules)  to the dynamics of the circadian clock and sleep. 

In the sleep and circadian modeling literature,  the newer physiology-based model shows better accuracy in predicting the alertness of shift workers than the phenomenology-based model, but the dynamics of physiological-based model have differential equations with different time scales, which pose challenges in optimization. To overcome the challenge, we propose a hybrid version of the PR model by applying singular perturbation techniques to reduce the system to a non-stiff, differentiable hybrid system. This reformulation facilitates the application of the calculus of variation and the gradient descent method to find the optimal light and sleep schedules that maximize the subjective alertness of shift worker. Our approach is validated through numerical simulations, and the simulation results demonstrate improved alertness compared to other existing schedules.

\section{Introduction}
Sleep is an important routine of our daily life. During sleep, our brain removes metabolic waste \cite{iliff2012paravascular,iliff2013brain} and supports the formation of long-term memories \cite{born2012system}. In preindustrial society, human routines were mainly determined by the natural light-dark cycle. In modern society, our habits are changed by the availability of artificial lighting, and shift work has become an essential part of our society that requires 24-hour service. However, sleep deprivation leads to increased sleepiness among shift workers, which in turn results in accidents and work-related injuries\cite{slater2012excessive,james2017shift}. This can have detrimental consequences not only for the shift workers themselves but also for those around them, especially in fields such as healthcare, transportation, and public safety, where shift work is common. Therefore, there is a critical need to address sleepiness during shift work and increase alertness of shift workers. The misalignments of circadian rhythms and enforced sleep-wake schedules are the main sources of increased sleepiness during shift work \cite{phillips2007quantitative,boivin2007working}. Adjusting circadian rhythms can be regarded as a control problem of a system with nonlinear dynamics\cite{doyle2006circadian}. The input of the system can be in the form of light, chemicals, and sleep schedules\cite{serkh2014optimal,abel2018controlling,booth2017one}. 
Optimal control theory provides ways to find the control inputs for a dynamic system such that the alertness of shift workers is maximized \cite{zhang2012optimal,yin2022human}. To find the optimal control input, we need mathematical models of circadian rhythms and sleep-wake cycles.
%explain what hybrid means in the introduction

 Many sleep-wake cycle models are derived from the Borbely’s phenomenology-based two-process model \cite{borbely1982two}. The transition between sleep and wakefulness is regulated by two processes: homeostatic sleep pressure, which increases during wakefulness and decreases during sleep, and circadian pacemaker, which is mainly affected by the light received by the retina. Achermann formulated a three-process model that combines the two-process model and sleep inertia \cite{achermann1994simulation,achermann1996time}. The three-process model was used to predict cognitive function and alertness \cite{reifman2004commentary, pettersson2019saccadic}. Yin et al. used the three-process model to calculate the light and sleep schedules as control input to optimize the alertness of workers  \cite{yin2022human}. However, as the physiology underlying sleep-wake dynamics is better understood, newer models that incorporate neuronal potential-level interactions are developed \cite{diniz2008delayed,phillips2007quantitative,rempe2010mathematical}. Although the three-process phenomenology-based model exhibits qualitative trends that match the data from the sleep deprivation study, this model lacks interpretation of physiological aspects. The Philips-Robinson (PR) model is a widely accepted physiology-based model\cite{phillips2010probing}. It associates the phenomenological concepts of sleep deprivations with the physiological mechanisms of sleep using a model that involves the potentials of two brain stems: the VLPO group, which is sleep-active, and the MA group, which is wake-active. The inclusion of physiological mechanisms makes the model sufficiently flexible to describe a wide range of phenomena, including alertness.
%introduce the optimization problem, then talk about the work of Jiawei, then about the limitation of three process model, then stiff ode
We perform numerical simulations of the three-process model and the PR model and compare the prediction of alertness with experimental data. We observe that the PR model can improve the prediction of subjective alertness over the three-process model. %these models are ode models. The odes are stiff. It is numerically hard to solve. Systems that represented by stiff odes are hard to simulate
However, the original PR model contains stiff differential equations, which means that the processes change on very different time scales. As recognized by Philips et al. \cite{phillips2007quantitative}, changes in neuronal potentials occur much faster than sleep homeostasis or circadian rhythms. 

The stiff equations of system dynamics pose challenges in optimization problems. %optimization problem is not clear, connect to 1st paragraph
The first challenge is the computational cost. Numerical solvers for stiff systems often require small time steps to maintain stability and accuracy. However, on a large time scale of multiple days, the computational costs increase significantly. The second challenge is the calculation of the gradient. For optimization techniques relying on gradients (e.g. gradient descent), the stiff equations complicate the computation of accurate and smooth derivatives. A recent work by Papatsimpa and Linnartz seeks to find the optimal light schedule that minimizes the difference between natural wake time and target wake time with the PR model \cite{papatsimpa2020personalized}, but the optimality of the greedy algorithm in the paper has not been proven. Hong et al. applied the PR model to analyze the sleep states of 21 shift workers and create a personalized plan designed to reduce the sleepiness of shift workers \cite{hong2021personalized}. Song et al. incorporate the sleep schedule in \cite{hong2021personalized} and propose a new schedule that aims to maximize the alertness of shift workers. However, these studies are based on empirical data rather than mathematical optimization.
In order to use the PR model to find optimal control inputs that maximize the alertness of shift workers, we propose a hybrid version of the PR model that has a hybrid system. We apply singular perturbation techniques (see Chapter 11 in \cite{khalil2002nonlinear}). The system is reduced to a hybrid system that exhibits both continuous and discrete dynamic behaviors. The discrete sleep state determines the mode of the system and the reduced model has continuous dynamics. The proposed hybrid PR model is non-stiff and differentiable so that we can use calculus of variation techniques to obtain the optimal solution. 
In this paper, we use the PR model to study the problem of optimizing the alertness of a shift worker using light and sleep schedules. We also compare our optimal schedules with existing sleep schedule in \cite{hong2021personalized} and demonstrate that our method can improve alertness in various scenarios.

%The method of using light to regulate circadian rhythm has been widely studied \cite{booth2017one,serkh2014optimal,yin2021optimization}. Yin et al. optimize light exposure and sleep schedules to minimize fatigue and cognitive impairment of shift workers \cite{yin2022human}. 

%very important
The main contributions of this paper are:
\begin{itemize}
    \item We propose a hybrid PR model that is non-stiff and differentiable while keeping the accuracy of alertness prediction of shift workers. 
    \item We extend the optimization method in \cite{yin2022human} with the hybrid PR model to optimize the alertness of shift workers using light and sleep schedules as input.
    \item We compare our optimal schedules with other existing schedules that are based on empirical data and experimental studies. Through numerical simulations, we demonstrate that our schedules can further improve alertness during shift work or during the full optimization horizon. 
    % other papers have empirical data and conducted experiments, I improve on them

\end{itemize}

\section{Model Derivation and Validation}% Make these seperated parts
%What is important about the PR model. Talk about comparison to 3 process model. Why the first point is interesting.
%Some histroy of the modeling, the meaning of PR model. The motivation of using PR model. This PR model has better alignment with experimental data. If we do optimal control with PR model, the system is stiff. 
%The hybrid model is better than 3 process. Then we look at two optimal problems.
\subsection{Existing Models}
\subsubsection{Three-Process Model}
Achermann developed a widely recognized three-process sleep regulation model, comprising three key components: the circadian pacemaker (Process C), the sleep homeostasis (Process S), and sleep inertia (Process W)\cite{achermann1994simulation}. The newer model updates components of the three-process model to incorporate newer discoveries \cite{yin2022human}.

\noindent {\bf Circadian pacemaker component:} The circadian pacemaker is represented by the Jewett-Forger-Kronauer (JFK) model \cite{jewett1999revised}. The states $x(t)$ and $x_c(t)$ are states of the core body temperature oscillator. The state $n(t)$ is the state of the light receptor in the retina. They follow the dynamics:
\begin{align}
\frac{dx}{dt} &= \frac{\pi}{12} [x_c + \mu(\frac{1}{3}x+\frac{4}{3}x^3-\frac{256}{105}x^7)+(1-0.4x)(1-k_c x_c) u],\label{eq:x}\\
\frac{dx_c}{dt} &= \frac{\pi}{12} [ - (\frac{24}{0.99729 \tau})^2 x + (q x_c - k x) (1-0.4x)(1-k_c x_c) u],\label{eq:xc}\\
u &= G \alpha_0 (\frac{I}{I_0})^p (1-\beta) (1-n),\label{eq:u}\\
\frac{dn}{dt} &= 60 [\alpha_0 (\frac{I}{I_0})^p (1-\beta) (1-n) -\gamma n],
\label{eq:Kronauer}
\end{align}
where all the parameters are listed in Table \ref{Table:TP_Parameter_Notation}.

\noindent {\bf Sleep Homeostasis component:} The sleep homeostasis represents the accumulation of sleep pressure that forms the sleep drive \cite{borb1999sleep}. The Process S has state $H(t)$ and it follows the dynamics:
\begin{align}
    \frac{dH}{dt} = \begin{cases} 
    -H/\tau_d,&\beta(t) = 1,\\
    (1-H)/\tau_r,&\beta(t) = 0,\end{cases}
\end{align}
where $\beta(t) = 1$ represents the subject is asleep and $\beta(t) = 0$ represents the subject is awake and $\tau_d = 4.2 h, \tau_r = 18.2 h$ are time constants.

\noindent {\bf Sleep Inertia component:} The sleep inertia represents a decrease in the level of mental alertness immediately after awakening \cite{achermann1994simulation,achermann1996time}. The process W has dynamics:
\begin{align}
    \frac{dW}{dt} = -\frac{1-\beta}{\tau_W}W,
\end{align}
where the time constant $\tau_W = 0.662h$. 

The sleep wake transition is governed by sleep propensity defined as:
\begin{align}
    \Phi(t) = H(t) - A_c x(t),
\end{align}
where $A_c = 0.1333.$ When $\Phi$ reaches the lower threshold $L_m$, the subject wakes up spontaneously. Likewise, when $\Phi$ reaches the upper threshold $H_m$, the subject falls asleep spontaneously. Sleep state $\beta(t)$ follows the discrete dynamics:
\begin{align}
    \beta(t)=F_{\beta}(t,\beta(t^-))=\begin{cases}
        1, &\Phi(t) = H_m,\\
        0, &\Phi(t) = L_m,\\
        \beta(t^-), &\textrm{otherwise},
    \end{cases}
    \label{eq:TP-cond}
\end{align}
where $\beta(t^-)=\lim_{\tau \xrightarrow{}t^-}\beta(\tau)$, $t^-$ represents the time just before $t$, and constants $H_m = 0.67, L_m = 0.17.$
The alertness is defined as
\begin{align}
    A_{TP}(t) = [1-\beta(t)][1+A_c x(t) - H(t) - W(t)],
\end{align}
and sleepiness level is defined as
\begin{equation}
    B(t) = 1 - A(t).
\end{equation}

\begin{table}[h]
\caption{\textbf{Three-process models}}%
\label{Table:TP_Parameter_Notation}
\centering
\begin{tabular}{c|c||c|c}
    Parameter & Value & Parameter & Value \\
    \hline
    $\mu$ & $0.13 h^{-1}$ & $k_c$ & $0.4h$\\
    \hline
    $q$ & $1/3$ & $\tau_x$ & $24.2h$\\
    \hline
    $k$ & $0.55h^{-1}$ & $G$& $33.75$\\
    \hline
    $\alpha_0$ & $0.05 h^{-1}$ &$p$ & $0.5$\\
    \hline
    $I_0$ & $9500 \textrm{ lux}$ & $\gamma$ & $0.0075h^{-1}$\\
    \hline
    $\tau_d$ & $4.2 h$ & $\tau_r$ & $18.2h$\\
    \hline
    $\tau_W$ & $0.662h$ & $A_c$ & $0.1333$ \\
    \hline
    $H_m$ & $0.67$ & $L_m$ & $0.17$ \\
    \hline
\end{tabular}
\end{table}

\begin{figure}[!htb]
 	\centering
        \includegraphics[width=0.5\textwidth]{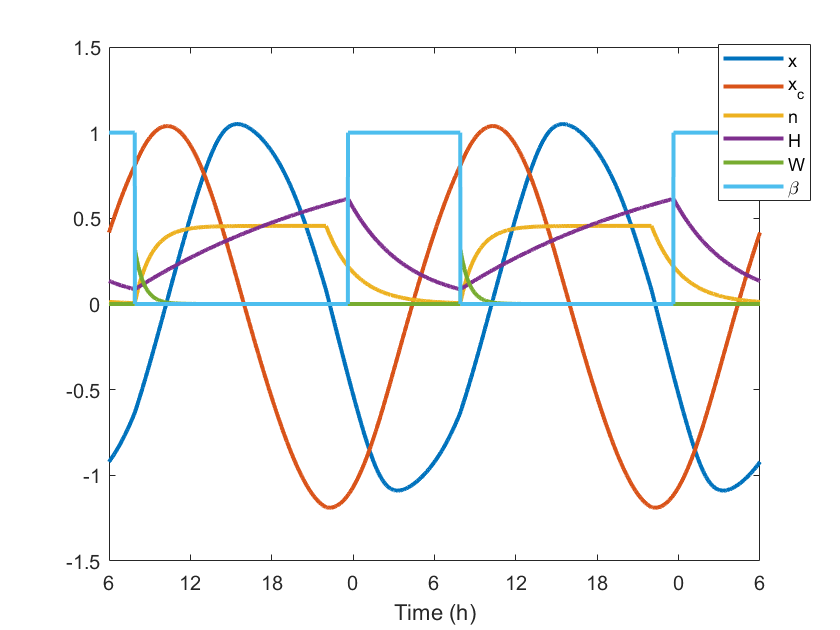}
	\caption{Periodic solution of three-process model accommodates to the periodic light schedule of 16-hour light 8-hour dark. 48 hours of the state trajectories are plotted here. }
	\label{fig:TP_per} 
\end{figure}

The states of the three-process model are denoted as
\begin{equation}
    \xi(t) = [x(t),x_c(t),n(t),H(t),W(t),\beta(t)]^T \in \mathbb{R}^6.
\end{equation}
We consider a 24-hour periodic light schedule as:
\begin{align}
    I_{ref}(t) = \begin{cases}
        150 \textrm{ lux}, &{\rm mod}(t,24) \in [0,16),\\
        0, &{\rm mod}(t,24) \in [16,24),
    \end{cases}\label{eq:light}
\end{align}
which simulate the indoor environment like hospital. The initial time $t = 0$ corresponds to 6 AM when the light is turned on and $t = 16$ corresponds to 10 PM when the light is turned off. The subject goes to sleep and wakes up spontaneously. At steady state, $\xi(t)$ forms a stable periodic solution, denoted as:
\begin{equation}
    \xi_{\rm ref}(t) = [x_{\rm ref}(t),x_{c,{\rm ref}}(t),n_{\rm ref}(t),H_{\rm ref}(t),W_{\rm ref}(t),\beta_{\rm ref}(t)]^T,
\end{equation}
in Fig. \ref{fig:TP_per}.

To simulate forced wakefulness, e.g., to study the effects of night shift work and sleep deprivation, we ignore the sleep-switching dynamics in Eq. \ref{eq:TP-cond} and force the sleep state $\beta(t) = 0$ during forced wakefulness.  

\subsubsection{Philips-Robinson (PR) Model} \label{section:PR model}
A widely used and more recently published sleep model is the Philips-Robinson (PR) model\cite{phillips2010probing}. This model consists of three parts: a circadian pacemaker component, a physiological component that models two mutually inhibitory neuron cell potentials, and a sleep homeostasis component.

\noindent {\bf Circadian pacemaker component:} The circadian pacemaker of the published PR model uses a circadian pacemaker model published in \cite{forger1999simpler}. In this paper, we update the PR model with the circadian pacemaker in \cite{jewett1999revised}. The circadian process has the same dynamics and parameters as Eq. \ref{eq:x}-\ref{eq:Kronauer} in three-process model. 

\noindent {\bf Physiological component:} The physiological components of the PR model are two mutually inhibitory neuron cell potentials. The monoaminergic (MA) potential that promotes wakefulness is represented by $V_m$ and the ventrolateral preoptic (VLPO) potential that promotes sleep is represented by $V_v$. The dynamics are defined by:
\begin{align}
    \frac{dV_m}{dt} = &\frac{1}{\tau_m} [-V_m -v_{mv} Q_v + A_m],\label{eq:dVm}\\
    \frac{dV_v}{dt} = &\frac{1}{\tau_v} [-V_v - v_{vm} Q_m + D_v],\label{eq:dVv}\\
    D_v &= - v_{vc} C + v_{vh} H + A_v, \label{eq:Dv}
\end{align}
where the firing rates $Q_m$ and $Q_v$ are sigmoidal functions of $V_m$ and $V_v$: 
\[ Q_m = \frac{Q_{max}}{1+{\rm exp}(-\frac{V_m-\theta}{\sigma})},\, Q_v = \frac{Q_{max}}{1+{\rm exp}(-\frac{V_v-\theta}{\sigma})}.
\] 
The term $D_v$ is the sleep drive and 
\[C = \frac{1}{2}(1+c_x x + c_{x_c} x_c)\] 
represents the circadian components. $V_m(t)$ is at a low level and $V_v(t)$ is at a high level during the sleep state, and vice versa during the wake state. 

\noindent {\bf Sleep homeostasis component:} The sleep homeostasis in PR model has different dynamics than the three-process model:
\begin{align}
    \frac{dH}{dt} = \frac{1}{\chi} (-H + \mu_H Q_m).
\end{align}
Sleep inertia is omitted in the PR model for two reasons: the effect is short-lived and the model is directly based on physiology, so the introduction of sleep inertia may lead to arbitrary fitting that degenerates the model.

All parameters of the PR model are listed in Table \ref{Table:PR_Parameter_Notation}. In \cite{postnova2018prediction}, Postnova et al. used a linear fitting procedure to align $C$ and $H$ with various cognitive performance metrics, using a distinct set of parameters for each metric. To avoid overfitting to a single metric, we use the definition of alertness in \cite{song2023real}:
\begin{align}
    A(t) = [1-\beta(t)][H^+(t)-H].
\end{align}

All the states of PR model can be denoted as
\begin{equation}
    \zeta(t) = [x(t),x_c(t),n(t),V_m(t),V_v(t),H(t)]^T \in \mathbb{R}^6
\end{equation}
Similar to three-process model, we apply the same periodic light schedule in Eq. \ref{eq:light}. The resulting periodic solution after entrainment are denoted as
\begin{equation}
    \zeta_{\rm ref}(t) = [x_{\rm ref}(t),x_{c,{\rm ref}}(t),n_{\rm ref}(t),V_{m,{\rm ref}}(t),V_{v,{\rm ref}}(t),H_{\rm ref}(t)]^T \in \mathbb{R}^6
\end{equation}

\begin{figure}
\begin{tabular}{cc}
  \includegraphics[width=60mm]{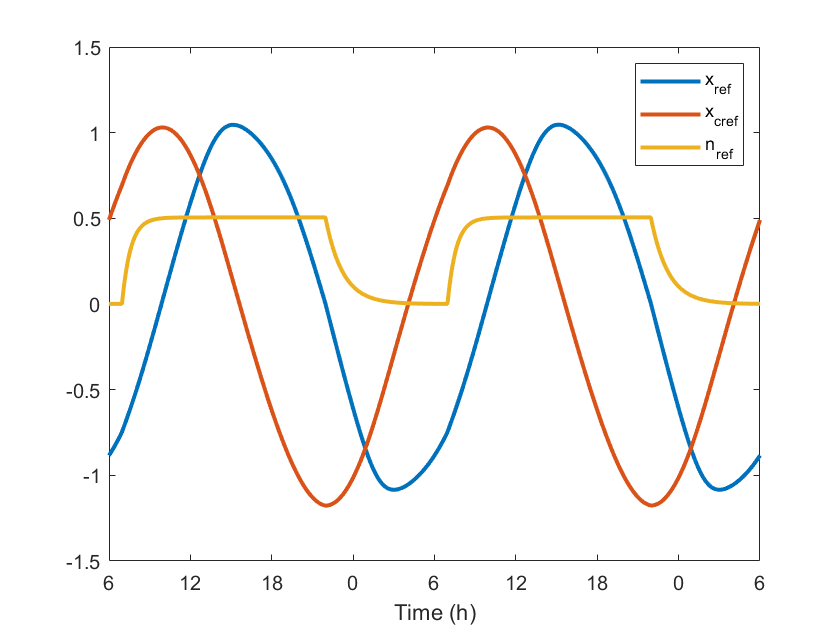} &   \includegraphics[width=60mm]{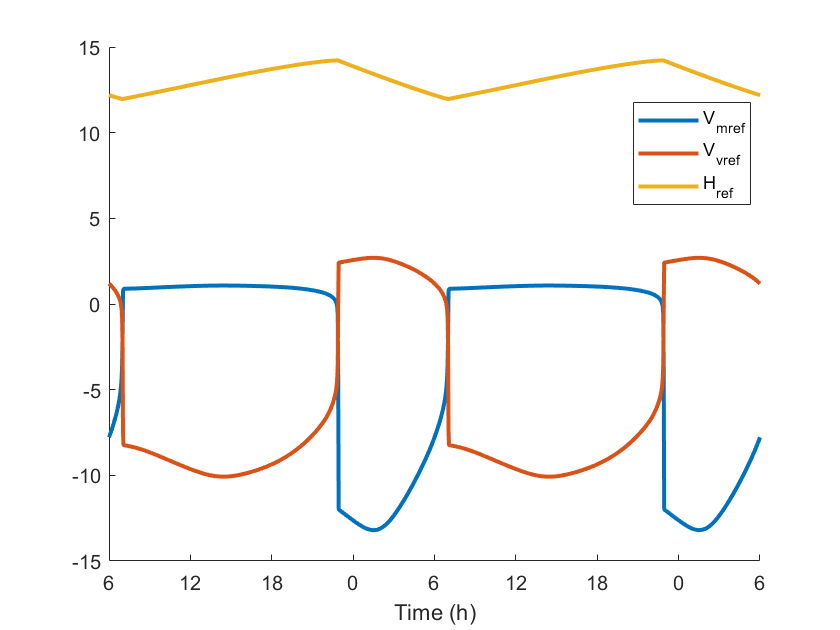} \\
(a) Circadian part & (b) MA,VLPO and sleep homeostasis \\[6pt]
\end{tabular}
\caption{Periodic solution of PR model accommodates to the periodic light schedule in Eq.\ref{eq:light}. Fig. 2(a) includes the circadian states $x,x_c,n$ and Fig. 2(b) includes the neuron potentials $V_m,V_v$ and sleep homeostasis $H$. }
\label{fig:PR Model}
\end{figure}
\begin{figure}[!htb]
 	\centering
        \includegraphics[width=0.5\textwidth]{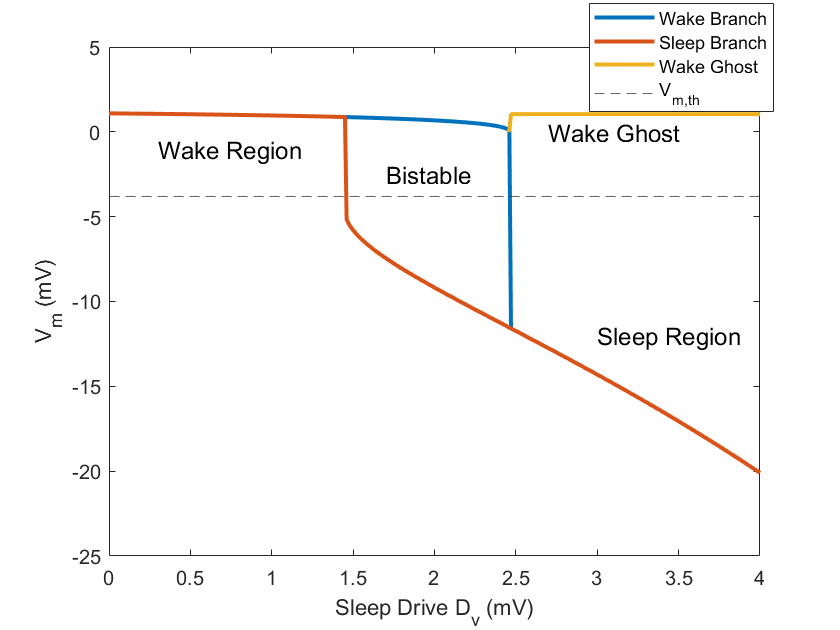}
	\caption{Bifurcation curves of PR model. The x-axis shows the sleep drive $D_v$ and y-axis shows the corresponding $V_m$ at equilibrium. In the wake and sleep region, the wake and sleep branch overlap because there is only one stable equilibrium. In the bistable branch, there are two stable equilibrium points for each $D_v$. When simulating sleep deprivation, we keep $V_m = 1.04 mV$.}
	\label{fig:bifur} 
\end{figure}

% the threshold Qm can be converted to a threshold in Vm, and then draw a line in figure 3
In the PR model, whether the subject is asleep or awake is determined by $V_m$. If $V_m$ is higher than a threshold $V_{m,{\rm th}} = -3.785 mV$ that corresponds to $Q_m > 1 s^{-1}$, the subject is awake. Since the time constants of the MA and VLPO dynamics in Eq.  \ref{eq:dVm}-\ref{eq:dVv} are much faster than the circadian clock and sleep homeostasis, we can use a quasi-static analysis and assume $D_v$ converges to the steady state equilibrium. The equilibrium point of $V_m$ as a function of $D_v$ can be obtained by setting the left-hand side of Eq. \ref{eq:dVm} and Eq. \ref{eq:dVv} to 0. The analytical form of $V_m$ as a function of $D_v$ is:
\begin{equation}
    V_m = - V_{mv} \cdot \left(\frac{Q_{max}}{1+{\rm exp} \left( \frac{ V_{vm}  \frac{Q_{max}}{1 + {\rm exp}( \frac{V_m - \theta}{\sigma})} - D_v - \theta }{\sigma} \right)}\right) + A_m \label{eq:DvVm}
\end{equation}
By analyzing the equilibrium points of $V_m$ for different $D_v$ in Eq. \ref{eq:dVm}-\ref{eq:dVv}, we have the bifurcation curves in Fig. \ref{fig:bifur}. Detailed analysis can be found in \cite{fulcher2008modeling}. If $D_v >2.46$, there is a unique equilibrium with low $V_m$ and therefore the subject is asleep. If $D_v <1.45$, there is a unique equilibrium with high $V_m$ and the subject is awake. If $1.45 \leq D_v  \leq 2.46$, there are two equilibria. The subject can be awake or asleep. This region is known as the bistable region. From the definition of $D_v$ in Eq. \ref{eq:Dv}, the sleep and wake threshold of $H$ can be derived:
\begin{align}
    H^+ = \frac{2.46-A_v+v_{vc} C}{v_{vh}},\\
    H^- = \frac{1.45-A_v+v_{vc} C}{v_{vh}}.
\end{align}
%we do not need sleep state, the full PR model is continuous, sleep just helps interpretation

 To simulate forced wakefulness in the PR model, a drive to $V_m$ is needed to maintain the wake state, known as \textit{wake effort} in \cite{fulcher2008modeling,fulcher2010quantitative,postnova2014effects}. Wake effort is an external stimulus that can take the form of a pharmacological agent or sensory input. With an additional wake effort to $V_m$, the subject will stay in the \textit{wake ghost} due to the characteristics of the system dynamics, keeping the subject awake. $V_m$ will be set to the maximum of $V_{m {\rm ,ref}}$ during periodic solution during forcedwakefulness.
 If $H < H^+$, no wake effort is needed to make the subject stay awake. If $H \geq H^+$ and the subject is required to maintain wake state, we keep $V_m$ at a fixed value, as the yelllow curve in Fig. \ref{fig:bifur}. % connects to the figure, showing highest of Vm ref
 Forced sleep is also reported in \cite{hong2021personalized,song2023real}. However, it is generally difficult to induce immediate sleep in an individual. Therefore, we do not consider the case of forced sleep.

The main differences between the three-process model and the PR model are summarized in Table \ref{Table:Model Comparison} in the next subsection.

\begin{table}[h]
\caption{\textbf{PR models}}%
\label{Table:PR_Parameter_Notation}
\centering
\begin{tabular}{c|c||c|c}
    Parameter & Value & Parameter & Value \\
    \hline
    $\mu$ & $0.13 h^{-1}$ & $k_c$ & $0.4h$\\
    \hline
    $q$ & $1/3$ & $\tau_x$ & $24.2h$\\
    \hline
    $k$ & $0.55h^{-1}$ & $G$& $33.75$\\
    \hline
    $\alpha_0$ & $0.05 h^{-1}$ &$p$ & $0.5$\\
    \hline
    $I_0$ & $9500 \textrm{ lux}$ & $\gamma$ & $0.0075h^{-1}$\\
    \hline
    $\tau_m$ & $1/360 h$ & $v_{mv}$ & $1.8 mVs$\\
    \hline
    $A_m$ & $1.3 mV$ & $\tau_v$ & $1/360 h$ \\
    \hline
    $v_{vm}$ & $2.1 mVs$ & $v_{vc}$ & $3.37 mVs$ \\
    \hline
    $v_{vh}$ & $1.01 mVnM^{-1}$ & $A_v$ & $-10.2 mV$ \\
    \hline
    $Q_{max}$ & $100 s^{-1}$ & $\theta$ & $10 mV$ \\
    \hline
    $V_{m,{\rm th}}$ & $-3.785 mV$ & $\sigma$ & $3 mV$ \\
    \hline
    $c_{x}$ & $0.8$ & $c_{x_c}$ & $-0.16$ \\
    \hline
    $\mu_H$ & $4.2 nMs$ & $\chi$ & $45 h$ \\
    \hline
\end{tabular}
\end{table}

    \subsection{Proposed Model}
%address the conversion of hybrid system with singular perturbation

\begin{figure}
\begin{tabular}{cc}
  \includegraphics[width=60mm]{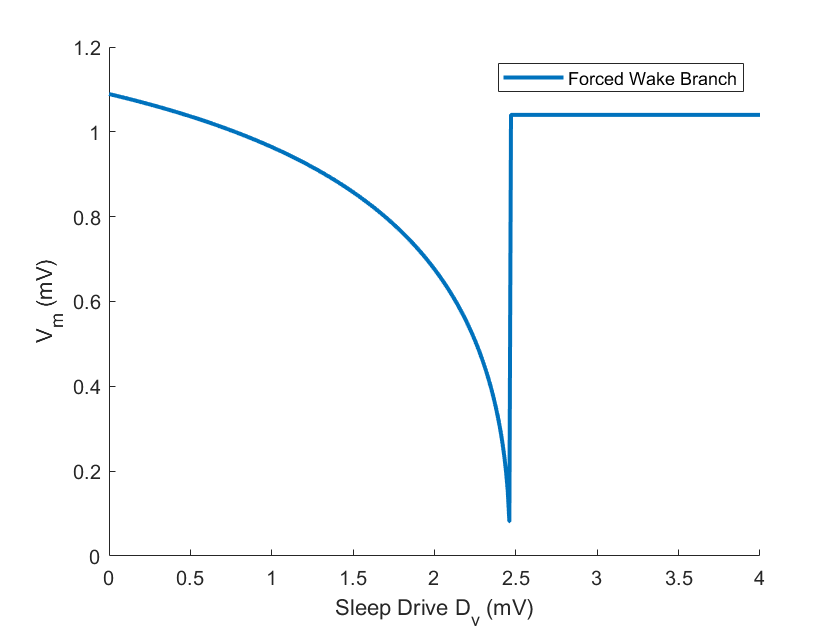} &   \includegraphics[width=60mm]{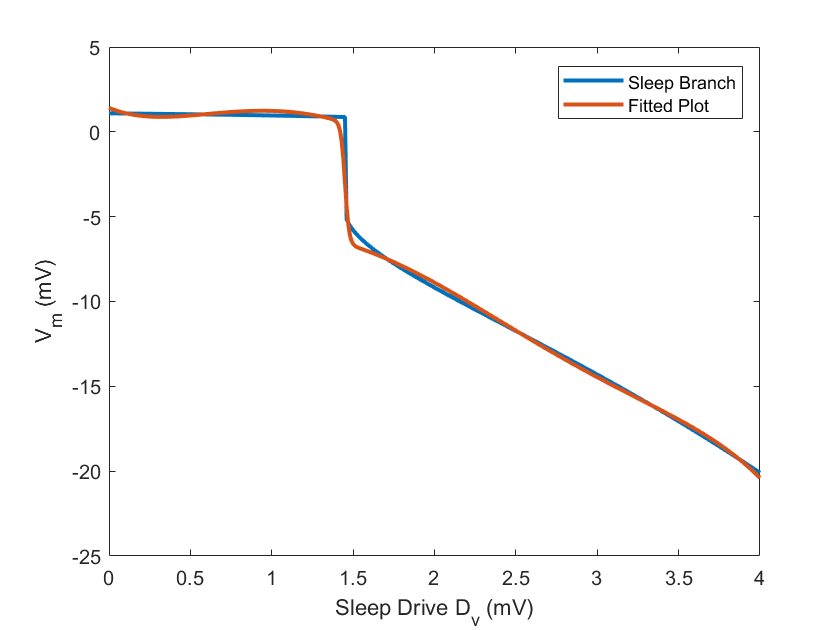} \\
(a) Forced Wake Branch & (b) Sleep Branch Fitting\\
\includegraphics[width=60mm]{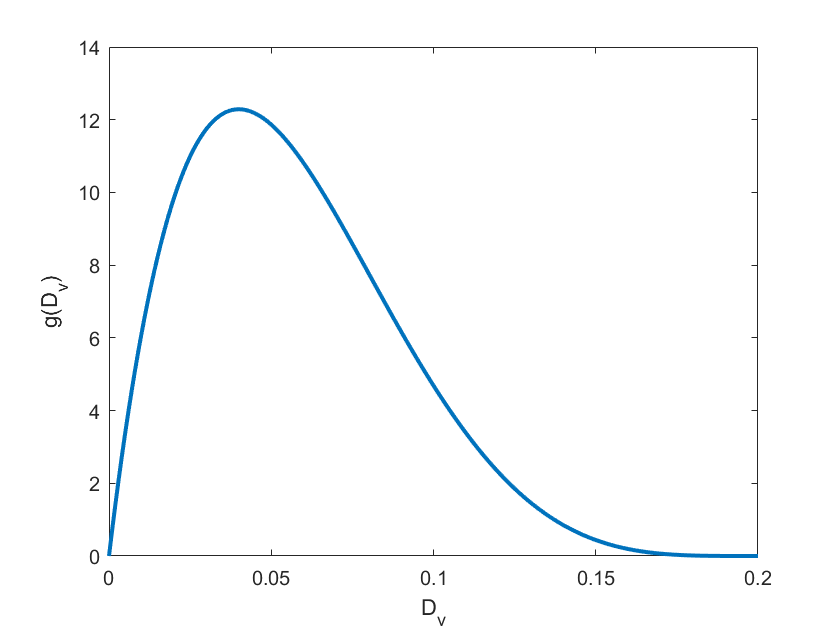} &   \includegraphics[width=60mm]{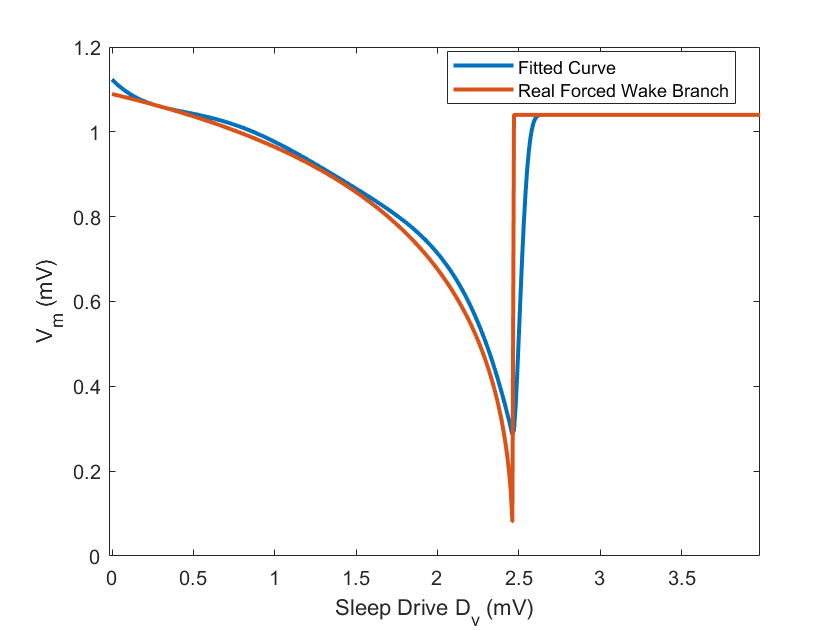} \\
(c) Filter $g(D_v)$ & (d)  Forced Wake Branch Fitting \\
\end{tabular}
\caption{(a) Forced wake branch. (b) Comparison of the real sleep branch (blue) and the fitted sleep branch in hybrid PR model (red) (c) The filter function $g(D_v)$ between $0 \leq D_v \leq 0.2$. (d) Comparison of the real forced wake branch (red) and the fitted forced wake branch in hybrid PR model(blue).}
\label{fig:hybrid PR Model}
\end{figure}

 To address the issues with the PR model, we propose a new version of the PR model that has hybrid behavior. To distinguish the difference between these two models, we refer to the original PR model as \textit{full PR model} and the hybrid version as \textit{hybrid PR model}. We use singular perturbation to reduce the model. The slow timescale components, i.e. circadian rhythms and sleep homeostasis, are still modeled by the continuous dynamics in the JFK model. The fast timescale components, such as $V_m$ and $V_v$, are assumed to reach steady states. The value of $V_m$ is derived from Eq. \ref{eq:DvVm} as a function of sleep drive $D_v$: $V_m = F(D_v)$ and $V_v$ is neglected. Similar to the three-process model, the sleep state $\beta(t)$ is discrete and change state spontaneously with the following dynamics:
\begin{align}
    \beta(t)=F_{\beta}(t,\beta(t^-))=\begin{cases}
        1, H(t) = H^+,\\
        0, H(t) = H^-,\\
        \beta(t^-), \textrm{otherwise},
    \end{cases}\label{eq:PR sleep dynamics}
\end{align}
 Depending on which of the three regions $D_v$ falls in, $V_m$ can be determined using the bifurcation analysis in the previous subsection. For the sleep or wake region, $V_m$ can be uniquely determined by Eq. \ref{eq:DvVm}. For the bistable region, $V_m$ is the larger root of Eq. \ref{eq:DvVm} if the subject is awake and the smaller root of Eq. \ref{eq:DvVm} if the subject is asleep. When the subject is in sleep deprivation, then $V_m =  \max(V_{m,{\rm ref}}).$ 
We can combine the wake region, the bistable region of the wake branch, and the wake ghost to the forced wake branch, shown in Fig. \ref{fig:hybrid PR Model} a. The subject will always be awake on the forced wake branch. With the sleep branch and the forced wake branch, we can determine the values of $V_m$ using $D_v$. 

In order to apply gradient-based optimization methods, such as the calculus of variation, the gradient of the optimization objective needs to be differentiable. We use two differentiable functions to model the sleep branch and the forced wake branch. We use the sum of a sigmoid function and a fifth-order polynomial to fit the sleep branch in Fig. \ref{fig:hybrid PR Model} b.
\begin{align}
    V_m = V_1(D_v) = 3.5702 \frac{\exp(-40 (D_v -1.45))-\exp(40 (D_v-1.45))}{\exp(-40 (D_v -1.45))+\exp(40 (D_v-1.45))} + \sum_{i = 0}^5 a_i D_v^i
\end{align}
Here $a_i$ are the coefficients of the polynomials and are listed in Table \ref{Table:Two Branches}. 
The coefficient values are obtained using the polyfit function in Matlab.

The forced wake branch is discontinuous at $D_v = 2.46.$ To make it smooth, we convolve the forced wake branch with a function in Fig. \ref{fig:hybrid PR Model}.c defined as follows:
\begin{align}
    g(D_v) = \begin{cases}
        0, &D_v \leq 0,\\
        468750 D_v (0.2-D_v)^4, &0 < D_v < 0.2,\\
        0, &D_v \geq 0.2.
    \end{cases}
\end{align}
The filter is continuous and the integration $\int_{-\infty}^{\infty} g(D_v) = 1$

After the convolution, the forced wake branch is approximated as 
\begin{align}
    V_m = V_2(D_v) = \begin{cases}
        \sum_{i = 0}^{5} b_i D_v^i , &D_v \leq 2.46,\\
        \sum_{i = 0}^{11} c_{i} D_v^i, &2.46 < D_v \leq 2.66,\\
        1.04, & D_v > 2.66.
    \end{cases}
\end{align}
Here $b_i$ and $c_i$ are all coefficients and are listed in Table \ref{Table:Two Branches}. 

\begin{table}[h]
\caption{\textbf{Parameters for Sleep branch model and Forced Wake Branch Model}}%
\label{Table:Two Branches}
\centering
\begin{tabular}{c|c||c|c}
    Parameter & Value & Parameter & Value \\
    \hline
    $a_0$ & $-3.2369$ & $a_1$ & $-3.9232$\\
    \hline
    $a_2$ & $9.2384$ & $a_3$ & $-7.3438$\\
    \hline
    $a_4$ & $2.0482$ & $a_5$& $-0.1964$\\
    \hline
    $b_0$ & $1.1236$ &$b_1$ & $-0.3960$\\
    \hline
    $b_2$ & $0.8783$ & $b_3$ & $-1.0640$\\
    \hline
    $b_4$ & $0.5328$ & $b_5$ & $-0.0982$\\
    \hline
    $c_0$ & $1.2155 \cdot 10^7$ & $c_1$ & $-2.5973 \cdot 10^7$ \\
    \hline
    $c_2$ & $2.2560 \cdot 10^7$ & $c_3$ & $-1.0007 \cdot 10^7$ \\
    \hline
    $c_4$ & $2.2781 \cdot 10^6$ & $c_5$ & $-2.0589 \cdot 10^5$ \\
    \hline
    $c_6$ & $-1.0268 \cdot 10^4$ & $c_7$ & $6.7223 \cdot 10^3$ \\
    \hline
    $c_8$ & $-3.4379 \cdot 10^3$ & $c_9$ & $1.2007 \cdot 10^3$ \\
    \hline
    $c_{10}$ & $-217.0005$ & $c_{11} $ & $16.6128$ \\
    \hline
\end{tabular}
\end{table}

The new hybrid system follows the dynamics:
\begin{align}
\frac{dx}{dt} &= \frac{\pi}{12} [x_c + \mu(\frac{1}{3}x+\frac{4}{3}x^3-\frac{256}{105}x^7)+(1-0.4x)(1-k_c x_c) u],\label{eq:x_PR}\\
\frac{dx_c}{dt} &= \frac{\pi}{12} [ - (\frac{24}{0.99729 \tau})^2 x + (q x_c - k x) (1-0.4x)(1-k_c x_c) u],\label{eq:xc_PR}\\
u &= G \alpha_0 (\frac{I}{I_0})^p (1-\beta) (1-n)\label{eq:u_PR},\\
\frac{dn}{dt} &= 60 [\alpha_0 (\frac{I}{I_0})^p (1-\beta) (1-n) -\gamma n],
\label{eq:Kronauer_PR}\\
\frac{dH}{dt} &= \frac{1}{\chi} \left(-H+\mu_H \frac{Q_{max}}{1+\exp(-\frac{V_m - \theta}{\sigma})}\right), \label{eq:H_PR}\\
V_m &= \begin{cases}
    V_1(D_v), \textrm{if } \beta = 1,\\
    V_2(D_v), \textrm{if } \beta = 0,
\end{cases}
\\
D_v &= - v_{vc} C + v_{vh} H + A_v \label{eq:Dv_PR}.
\end{align}
The model parameters in Eq. \ref{eq:x_PR}-\ref{eq:H_PR} and Eq. \ref{eq:Dv_PR} are same as the parameters in full PR model.
The alertness of hybrid PR model can also be defined as:
\begin{align}
    A(t) = [1-\beta(t)][H^+(t)-H].
\end{align}
The states of the hybrid PR models are denoted as
\begin{equation}
    X(t) = [x(t),x_c(t),n(t),H(t),\beta(t)]^T \in \mathbb{R}^5.
\end{equation}
When the light schedule is the same as Eq. \ref{eq:light}, and the sleep state $\beta(t)$ in the hybrid PR model follows Eq. \ref{eq:PR sleep dynamics}, we obtain the periodic solution
\begin{equation}
     X_{ref}(t) = [x_{ref}(t),x_{cref}(t),n_{ref}(t),H_{ref}(t),\beta_{ref}(t)]^T. \label{eq:hybrid PR ref}
\end{equation}

\begin{figure}
\begin{tabular}{cc}
  \includegraphics[width=60mm]{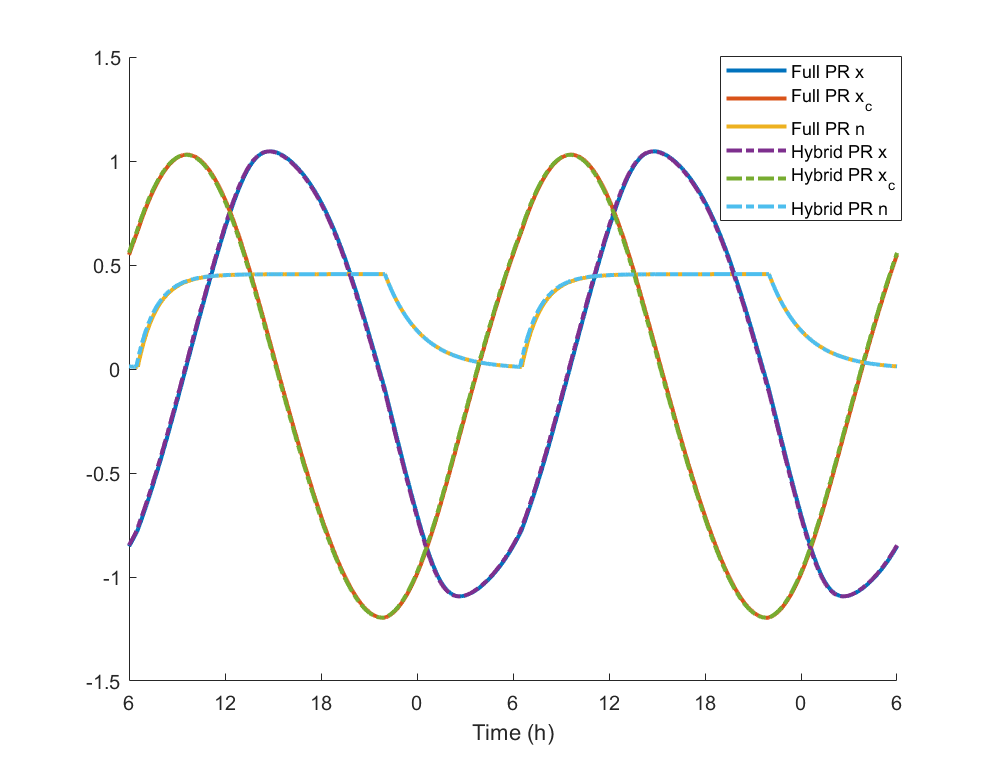} &   \includegraphics[width=60mm]{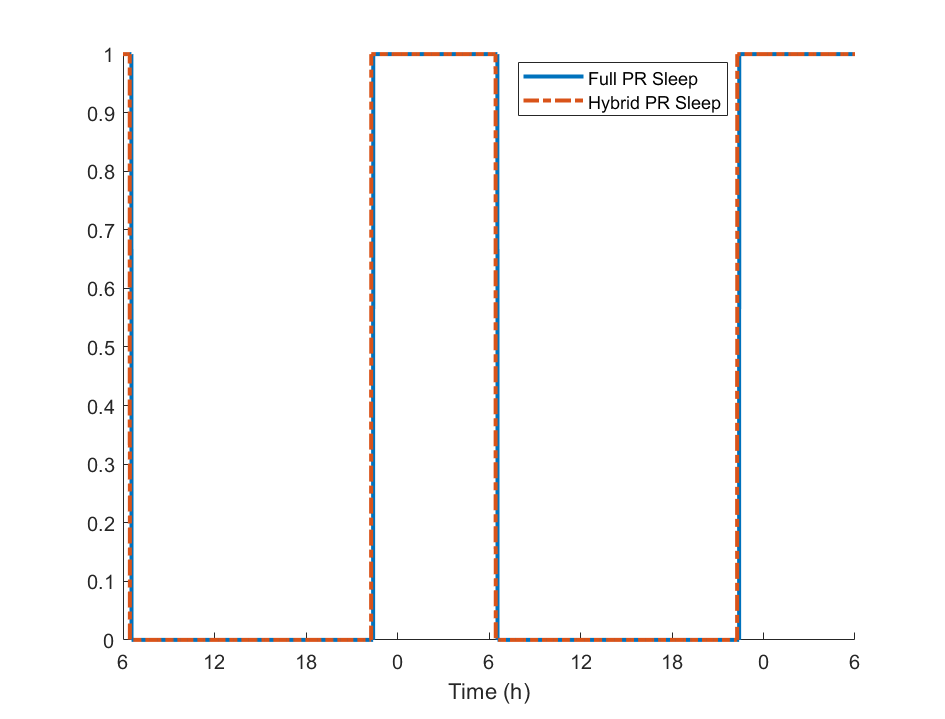}\\
(a) Comparison of Circadian Rhythms & (b) Comparison of Sleep Time \\
  \includegraphics[width=60mm]{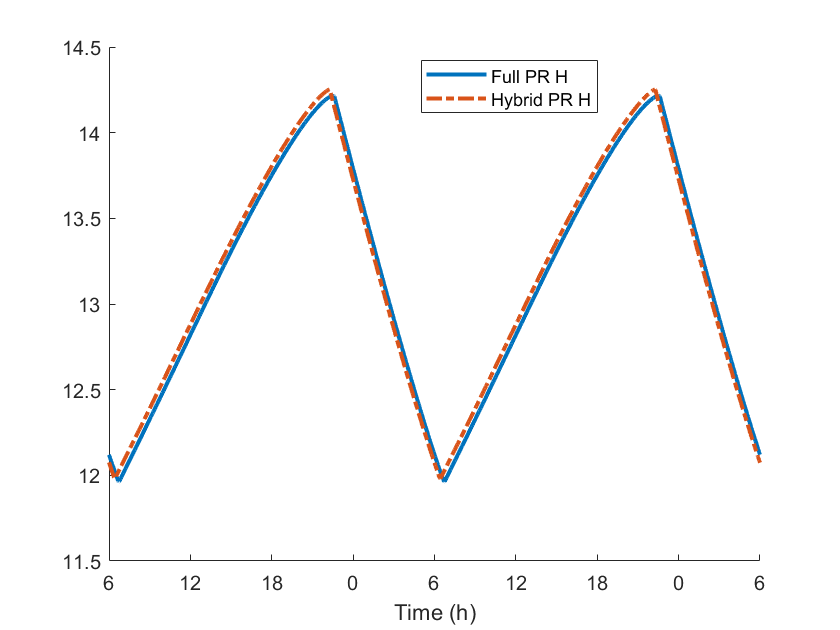}&   \includegraphics[width=60mm]{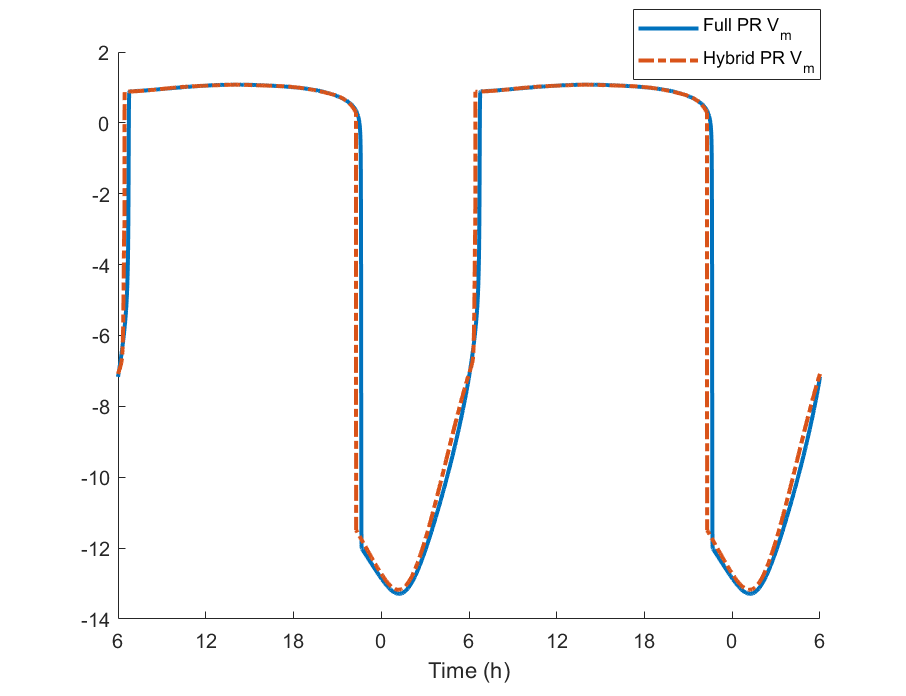}\\
(c) Comparison of $H$ & (d) Comparison of $V_m$
\end{tabular}
\caption{These are the comparisons of three models. The x axes show the time in hours. The starting times are 6 AM and the durations are 48 hours. (A) Circadian rhythm Process C which includes states $x,x_c$ and $n$. These three states have the identical dynamics in the full PR model and hybrid PR model. (b) Sleep state $\beta(t)$. 0 means the subject is awake and 1 means the subject is asleep. For the full PR model, the subject wakes up at 6:33 AM and sleeps at 10:25 PM. For the hybrid PR model, the subject wakes up at 6:27 AM and sleeps at 10:17 PM. (c) Comparison of the sleep homeostasis between the full PR model and hybrid PR model (d) Comparison of the $V_m$ between the full PR model and hybrid PR model.}\label{fig:PR periodic comp} 
\end{figure}

To demonstrate the accuracy of our approximation using the proposed hybrid PR model, we compare the state trajectories of the hybrid PR model to the full PR model. The periodic solutions of circadian rhythm $x_{ref},x_{cref}$ and $n_{ref}$ are plotted in Fig. \ref{fig:PR periodic comp}(a) and the sleep and wake-up time are plotted in Fig. \ref{fig:PR periodic comp}(b). The phase of circadian rhythm is defined as $-\tan(\frac{x_c}{x})$ and the average phase difference between the hybrid PR model and full PR model is 0.0126 radians, which is equivalent to 2.89 minutes. The hybrid model wakes up 6 minutes earlier and goes to sleep 8 minutes earlier than the full hybrid model. The states $H$ and $V_m$ are plotted in Fig. \ref{fig:PR periodic comp}(c) and Fig. \ref{fig:PR periodic comp}(d). We can see that the hybrid PR model is a good approximation of the full PR model.

\begin{table}[h]
\caption{\textbf{Comparison of Three-Process Model, PR Model and Hybrid PR model}}%
\label{Table:Model Comparison}
\centering
\begin{tabular}{c||c|c|c}
     & Three-process Model & PR model & Hybrid PR Model \\
    \hline
    Number of States & 6 & 6 & 5\\
    \hline
    System Dynamics & Hybrid & Continuous & Hybrid\\
    \hline
    Stiffness & Non-stiff & Stiff & Non-stiff\\
    \hline
\end{tabular}
\end{table}

\subsection{Validation}
To validate the precision of cognitive performance prediction, we perform two simulations for all three models discussed and fit the predicted alertness data to the experimental data in \cite{johnson1992short,wehr1992short}. 
In \textit{constant routine experiment protocol} \cite{johnson1992short}, subjects were instructed to remain awake for an extended duration of 60 hours under continuous light conditions (150 lux). Subjects access their subjective alertness by indicating their level of alertness on a linear 100-mm visual-analog scale (VAS) \cite{monk1987subjective}. The experimental data have the subjective alertness measurement at $n$ different time steps and the mean and covariance are denoted as 
\begin{equation}
    M_{VAS} = [\mu_{VAS1}, \cdots, \mu_{VASn}]^T
\end{equation} and 
\begin{equation}
\Sigma_{VAS} = \begin{bmatrix}
\sigma_{VAS1}^2 & 0 & \cdots & 0 \\
0 & \sigma_{VAS2}^2 & \cdots & 0 \\
\vdots & \vdots & \ddots & \vdots \\
0 & 0 & \cdots & \sigma_{VASn}^2
\end{bmatrix}.
\end{equation} We use linear fitting to find the parameters $\theta_{VAS} = [\theta_{VAS1},\theta_{VAS2}]$ that minimizes the normalized mean square error (NMSE): 
\begin{equation}
    NMSE_{VAS} = (M_{VAS}-(\theta_{VAS1} A + \theta_{VAS2} \mathbf{1}))^T \Sigma_{VAS}^{-1} (M_{VAS}-(\theta_{VAS1} A + \theta_{VAS2} \mathbf{1}))
\end{equation}
where $A=[A_1,A_2,\cdots,A_n]^T$ represents the predicted alertness at time step $1$ to $n$ and $\mathbf{1}\in \mathbb{R}^n$ is a vector of ones. 
The parameters and NMSEs for each model are listed in Table \ref{Table:Alertness Fitting Params}.

In \textit{photoperiod experiment} \cite{wehr1992short}, subjects were instructed to stay in 16 h daylight in one week and then stay in dim light less than 1 lux for 24 hours. The Stanford Sleepiness Scale \cite{shahid2011stanford} are recorded during the dim light period. Similarly, the experimental data have mean \begin{equation}M_{SSS} =[\mu_{SSS1},\mu_{SSS2},\cdots,\mu_{SSSn}]^T\end{equation} and \begin{equation}\Sigma_{SSS} = \begin{bmatrix}
\sigma_{SSS1}^2 & 0 & \cdots & 0 \\
0 & \sigma_{SSS2}^2 & \cdots & 0 \\
\vdots & \vdots & \ddots & \vdots \\
0 & 0 & \cdots & \sigma_{SSSn}^2
\end{bmatrix}.
\end{equation}
We also find the parameters $\theta_{SSS} = [\theta_{SSS1},\theta_{SSS2}]$ that minimizes the NMSE for SSS:
\begin{equation}
    NMSE_{SSS} = (M_{SSS}-(\theta_{SSS1} A + \theta_{SSS2} \mathbf{1}))^T \Sigma_{SSS}^{-1} (M_{SSS}-(\theta_{SSS1} A + \theta_{SSS2} \mathbf{1}))
\end{equation}
The parameters for all three models are listed in Table \ref{Table:Sleepiness Fitting Params}.

\begin{table}[h]
\caption{\textbf{Alertness Linear Fitting Parameters}}%
\label{Table:Alertness Fitting Params}
\centering
\begin{tabular}{c||c|c|c}
     & Three-process Model & PR model & Hybrid PR Model \\
    \hline
    $\theta_{VAS1}$ & 72.92 & 9.17 & 9.52\\
    \hline
    $\theta_{VAS2}$ & 22.19 & 53.25 & 55.19\\
    \hline
    $NMSE_{VAS}$ & 32.99 & 29.24 & 25.37\\
    \hline
\end{tabular}
\end{table}

\begin{table}[h]
\caption{\textbf{Sleepiness Linear Fitting Parameters}}%
\label{Table:Sleepiness Fitting Params}
\centering
\begin{tabular}{c||c|c|c}
     & Three-process Model & PR model & Hybrid PR Model \\
    \hline
    $\theta_{SSS1}$ & -0.656 & 3.99 & -0.656 \\
    \hline
    $\theta_{SSS2}$ & 2.72 & 0.057 & 2.78\\
    \hline
    $NMSE_{SSS}$ & 29.22 & 8.61 & 9.55\\
    \hline
\end{tabular}
\end{table}

\begin{figure}[!htb]
\begin{tabular}{p{60mm} p{60mm}}
  \includegraphics[width=60mm]{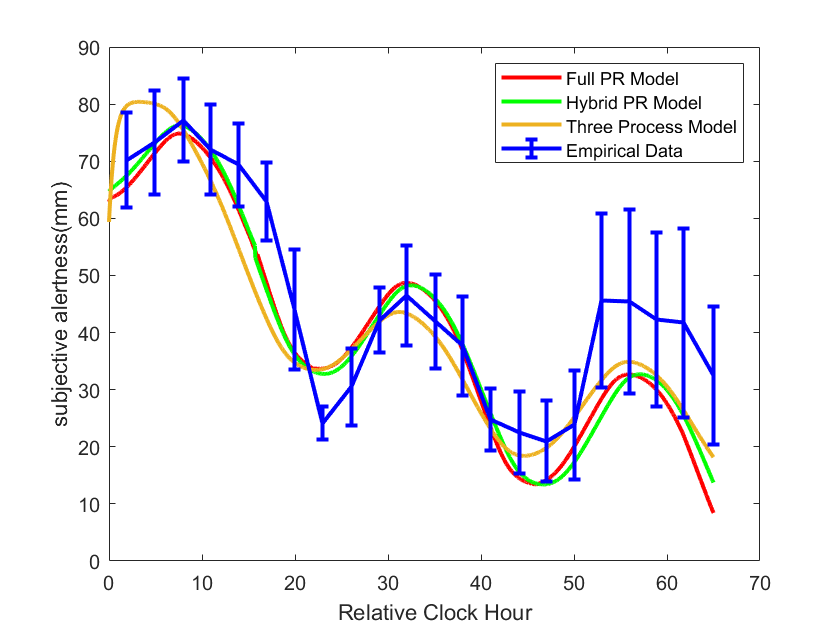} &   \includegraphics[width=60mm]{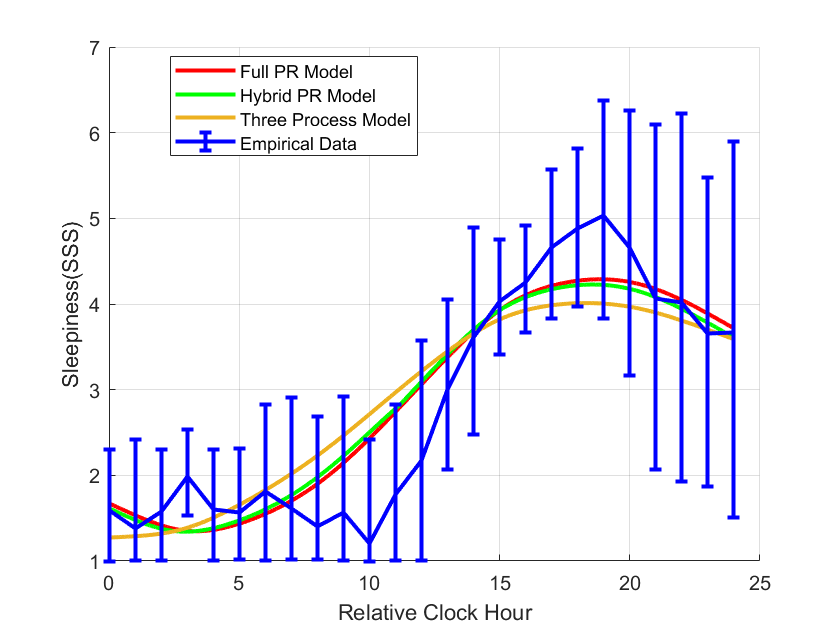}\\
(a) Alertness Fitting Comparison & (b) Sleepiness Fitting Comparison.
\end{tabular}
\caption{
Alertness and Sleepiness Fitting Comparison. The x-axis shows the time that subject stays awake since wake up time. The y-axis shows the subjective alertness or sleepiness. The empirical data shows the mean and plus minus standard deviation of the experimental data.}\label{fig:alertness sleepiness comp} 
\end{figure}

We plot the fitted subjective alertness and sleepiness in Fig. \ref{fig:alertness sleepiness comp}. The hybrid PR model has the lowest $NMSE_{VAS}$ among three models. The $NMSE_{SSS}$ of PR model is also very close to the full PR model and significantly lower than the three-process model. The alertness $A$ predicted by the hybrid PR model shows close agreement with the empirical data in \cite{johnson1992short,wehr1992short}.

%the title should be the objective:
% related to optimal alertness
\section{Optimal Light and Sleep Schedule For Improving Alertness}
Extensive research has aimed to enhance the alertness of shift workers by implementing structured sleep schedules\cite{aakerstedt1998there,vital2021optimal,song2023real}. For shift workers, the subject must remain awake during scheduled night shifts. In \cite{yin2022human}, Yin et al. proposed the problem of optimizing alertness with light and sleep schedules together with the three-process model. Inspired by \cite{yin2022human}, we discuss two problems for optimizing alertness with the hybrid PR model because the PR model can predict alertness more accurately than the three-process model.
For the first problem, we find the optimal light and sleep schedules during the optimization horizon to maximize the cumulative alertness only during the shift work.
For the second problem, we find the optimal light and sleep schedules during the optimization horizon that maximize the cumulative alertness during the full horizon. We optimize not only alertness during shift work but also during the time the subject is resting. The details of these problems will be defined in the following subsections.
In both problems, light is subject to the constraint:
\begin{equation}
    0 \leq I(t) \leq 150\textrm{ lux}.\label{eq:light con}
\end{equation}
We adopt the `tunable sleep schedule' in \cite{yin2021optimization}. The subject can stay awake when $H^+ \leq H(t) \leq H^+_{max}$ where $H^+_{max} = \frac{3.43-A_v + v_{vc}C}{v_{vh}}$ corresponds to $H$ staying awake for 2 hours after reaching the sleep threshold. The introduction of this upper limit prevents excessive sleep deprivation. The subject can also wake up earlier than the spontaneous wake time when $H^- \leq H(t) \leq  H^-_{max}$ where $H^-_{max} = \frac{2.11-A_v + v_{vc}C}{v_{vh}}$ corresponds to the value of $H$ 2 hours before reaching the wake threshold. We assume that the subject is always awake ($\beta(0)=0$) at the beginning of optimization. The sleep state during tunable sleep schedule can be expressed as
\begin{align}
    \beta(t) = F_{\beta}^{\prime}(t,\beta(t^-)) = \begin{cases}
        1, & t\in \{t_{sleep}^1, \cdots, t_{sleep}^{N_f}\},\\
        0, & t\in \{t_{wake}^1, \cdots, t_{wake}^{M_f}\},\\
        \beta(t^-), &\textrm{otherwise},
    \end{cases} \label{eq:tunable}
\end{align}
with the comfort constraints
\begin{equation}
    H^-\leq H(t_{sleep}^i) \leq H^-_{max}, H^+\leq H(t_{wake}^j) \leq H^+_{max}
    \label{eq:tunable constr}
\end{equation}
where $t_{sleep}^i$ and $t_{wake}^j$ represent the $i$-th sleeping time and $j$-th wake-up time. Here, $N_f$ and $M_f$ denote the number of times the subject falls asleep and wakes up, respectively, during the optimization period. $I(t),t_{sleep}^1, \cdots, t_{sleep}^{N_f}$, $t_{wake}^1, \cdots, t_{wake}^{M_f}$ are all optimization variables. Note that the shift work are predetermined by user's demand, so the sleep state during shift work will always be awake and does not need to satisfy the comfort constraints.

\subsection{Shift Work Alertness Optimization}\label{shift work alertness optimization}
In this subsection, we solve the shift work alertness optimization problem. The states of the subject $X$ follow the hybrid dynamics that is piecewise continuous and the dynamics switch between asleep mode ($\beta(t)=1$) and awake mode ($\beta(t)=0$).
Assume that the dynamics for the $i$-th mode is described by the following dynamics:
\begin{equation}
    \dot{X} = D_i(X,I), t\in[t_{i-1},t_i),\forall i \in\{1,2,\cdots,N\},\label{eq:hybrid}
\end{equation}
where $N$ is the total number of modes during the optimization time horizon. All switching times $t_i$ are subject to the inequality constraints in the tunable schedule in Eq. \ref{eq:tunable constr}, except during shift work. The shift work alertness optimization problem is:

Given the initial conditions $X(0) = X_0$, the dynamics of the hybrid model represented as $\dot{X} = D_i(X,I,\beta)$ for $i = 1,\cdots, N$, the light constraint in Eq. \ref{eq:light con} and tunable sleep schedule, we maximize the cumulative alertness of a subject during shift work over multiple days:
\begin{equation}
    \max \int_{t \in t_{work}} (1-\beta(t)) [H^+(t)-H(t)] dt,
\end{equation}
where $t_{work}$ represents the set of shift work intervals. 

We apply the variational calculus method in \cite{yin2022human} to solve the alertness optimization problem. 
An objective function is formulated as 
\begin{equation}
    J = \int_0^{t_f} L(t,X,I)dt, \label{eq:def J}
\end{equation}
and $L = - \mathbbm{1}_{work} [H^+(t)-H(t)]$, where 
\begin{equation}
    \mathbbm{1}_{work}(t) = \begin{cases}
        1, & \textrm{if } t \in t_{work},\\
        0, & \textrm{if } t \not\in t_{work},
    \end{cases}
\end{equation}
is an indicator function. We use the standard constrained optimization approach by introducing Lagrange multipliers $\lambda(t)\in \mathbf{R}^4$.
The augmented objective function is:
\begin{align}
    J_a = \int_0^{t_f} L(\tau,X,I) d\tau + \sum_{i=1}^{N} \int_{t_{i-1}}^{t_i} \lambda(\tau)[D_i(X,I)-\dot{X}(\tau)] d\tau.
\end{align}
The selection of $\lambda$ does not affect the value of $J_a$. By selecting the Lagrange multiplier $\lambda(t)$ as:
\begin{align}
    &\lambda(t_N) = 0,\label{eq:final costate}&\\
    \dot{\lambda}(t)=-\frac{\delta L(t,X,I)}{\delta X} - &\left(\frac{\delta D_i(X,I)}{\delta X}\right)^T \lambda(t) \textrm{, when } t\in [t_{i-1},t_i),\label{eq:costate}\\
    &\lambda(t_i^-) = \lambda(t_i^+),\label{eq:costate switch}
\end{align}
the first variation of augmented cost $\delta J_a$ simplify to:
\begin{align}
    \delta J_a = \sum_{i=1}^N \int_{t_{i-1}}^{t_i} \nabla_{I(t)} J \delta I d \tau + \sum_{i=1}^{N-1} \nabla_{t_i}J \delta t_i,\label{eq:grad I T}
\end{align}
\begin{equation}
    \nabla_{I(t)} J = \frac{\partial L(t,X,I)}{\partial I} + \lambda^T(t) \frac{\partial D_i(X,I)}{\partial I},\label{eq:grad I}
\end{equation}
\begin{equation}
\begin{split}
    \nabla_{t_i}J = &L(t_i^-,X(t_i^-),I(t_i^-))-L(t_i^+,X(t_i^+),I(t_i^+))+ \lambda(t_i^-) D_i(X(t_i^-),I(t_i^-)) %\label{eq:grad I}
    \\&- \lambda^T(t_i^+) D_{i+1}(X(t_i^+),I(t_i^+))\label{eq:grad t}
\end{split}
\end{equation}
where $\nabla_{I(t)} J$ and $\nabla_{t_i}J$ are the gradient of $J$ with respect to $I$ and $t_i$ and $\delta I, \delta t_i$ are perturbations in $I$ and $t_i$. The details of $\lambda$ and gradient formulations are in the Appendix.

Applying gradient descent method, at the $j$-th iteration, the light input $I^j(t)$ and i-th swtiching time $t_i^j$ can be updated by:
\begin{equation}
    I^{j+1}(t) = \min(\max(0,I^j(t) - \eta_I \nabla_{I(t)}J),I_{\max}),\label{eq:I grad update}
\end{equation}
\begin{equation}
    t^{j+1}_i = t_i^j - \eta_{t} \nabla_{t_i^j}J \in \Omega_{t_i},\label{eq:t grad update}
\end{equation}
where $\eta_I$ and $\eta_{t}$ are the step size for update and $\Omega_{t_i}$ represents the set of feasible $t_i$ that satisfy the constraint Eq. \ref{eq:tunable constr}.

The steps of the gradient descent algorithm can be summarized as follows.
\begin{tcolorbox}
[colback=blue!5!white,colframe=blue!75!black,title=Gradient Descent Algorithm]
\begin{description}
    \item[Step 1:] Start with an initial guess of light $I^0(t)$ and sleep schedule $t_i^0, i\in\{1,2,\cdots,N-1\}$ iteration $j = 0$;
    \item[Step 2:] Integrate the state dynamics with Eq. \ref{eq:hybrid} forward in time to obtain state trajectory $X^j(t)$ and the objective function $J^j$;
    \item[Step 3:] Integrate the Lagrange multiplier with Eq. \ref{eq:final costate} - \ref{eq:costate switch} backward in time to obtain $\lambda^j(t),t\in[t_0,t_f]$;
    \item[Step 4:] Calculate the gradients $\nabla_{I(t)^j}J$ and $\nabla_{t_i}^j$ with Eq. \ref{eq:grad I} and Eq. \ref{eq:grad t}, then perform updates to $I^{j+1}$ and $t_i^{j+1}$ with Eq. \ref{eq:I grad update}, \ref{eq:t grad update};
    \item[Step 5:] Repeat Step 2 to 4 until light and sleep schedules reach convergence.
\end{description}
\end{tcolorbox}

In Step 1 of the gradient descent algorithm, different initial guesses can lead to different local minima. We construct an initial guess based on periodic solution in Eq. \ref{eq:hybrid PR ref} with the following steps:
\begin{tcolorbox}[colback=blue!5!white,colframe=blue!75!black,title=Initial Guess Based On Periodic Solution]
\begin{description}
    \item[Step 1:] Start with light and sleep schedules that follow the periodic solution in Eq. \ref{eq:hybrid PR ref};
    \item[Step 2:] Change all sleep states during shift work to awake;
    \item[Step 3:] Simulate the system forward with the updated sleep schedule and obtain $N$ episodes of sleep so that the i-th episode $\{t^i|t_{sleep}^i \leq t^i < t_{wake}^i\}$ satisfies $\beta(t_i) = 1$, sleep state during shift work will remain awake;
    \item[Step 4:] Calculate the updated $H(t)$ with updated sleep in Step 3,
    adjust the begin and end time of sleep episode $i$, $t_{sleep}^i$ and $t_{wake}^i$ so that $t_{sleep}^i = \min\{\beta(t^i) == 1 \land H^+ \leq H(t^i) \leq H^+_{max}\}$, and $t_{sleep}^i = \max\{\beta(t^i) == 1 \land H^- \leq H(t^i) \leq H^-_{max}\}$, where $\land$ represents logic `and';
    \item[Step 5:] Increase $i$ by 1;
    \item[Step 6:] Repeat Step 3 to 5 until $i > N$, the number of sleep episodes.
\end{description}
\end{tcolorbox}

This initial guess will provide a sleep schedule that satisfies the comfort constraints in Eq. \ref{eq:tunable constr}. 

\begin{figure}
\begin{tabular}{p{60mm} p{60mm}}
  \includegraphics[width=60mm]{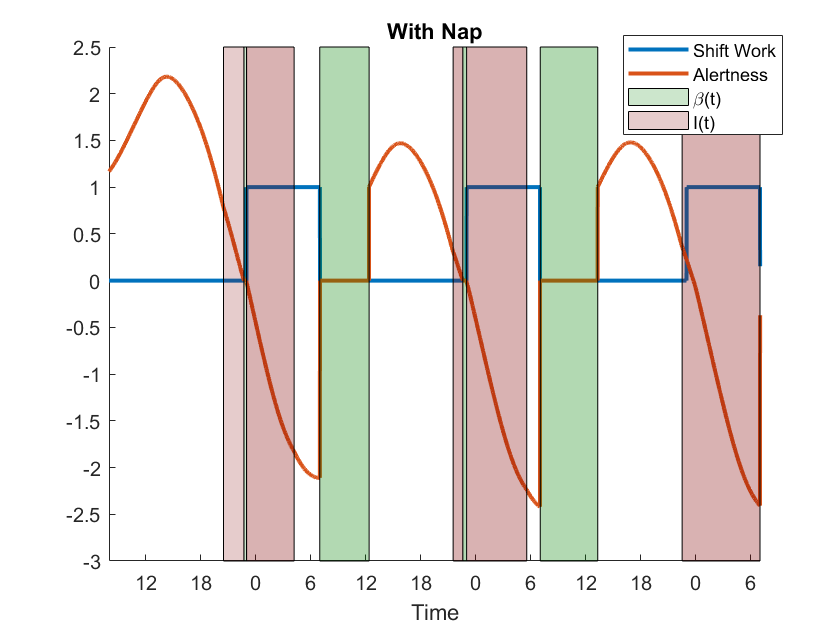} &   \includegraphics[width=60mm]{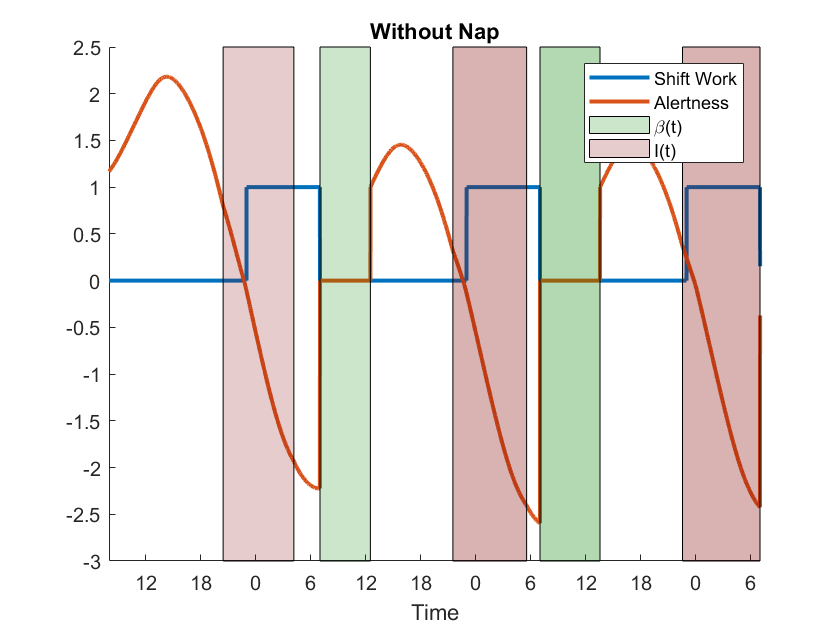}\\
(a) Optimal Schedules With Naps & (b) Optimal Schedules Without Naps \\
  \includegraphics[width=60mm]{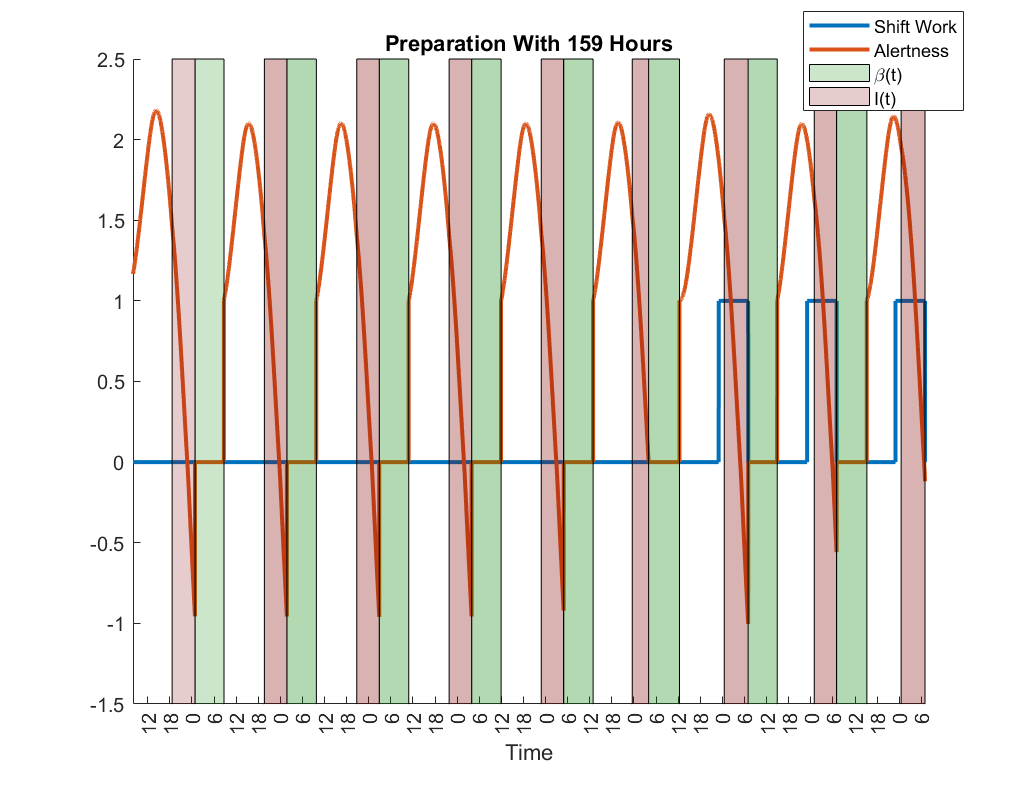}&   \includegraphics[width=60mm]{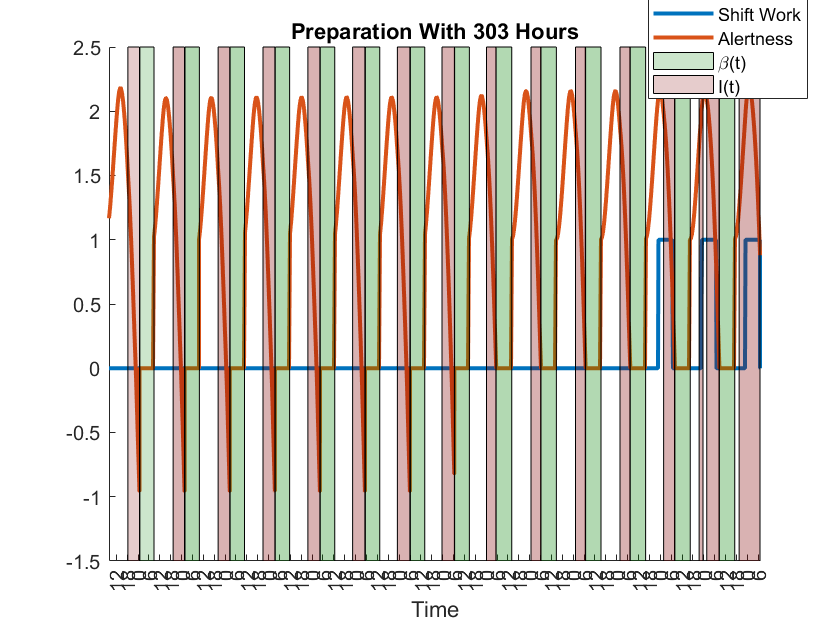}\\
(c) Optimal Schedules With 6 Days Preparation & (d) Optimal Schedules With 13 Days Preparation\\
\includegraphics[width=60mm]{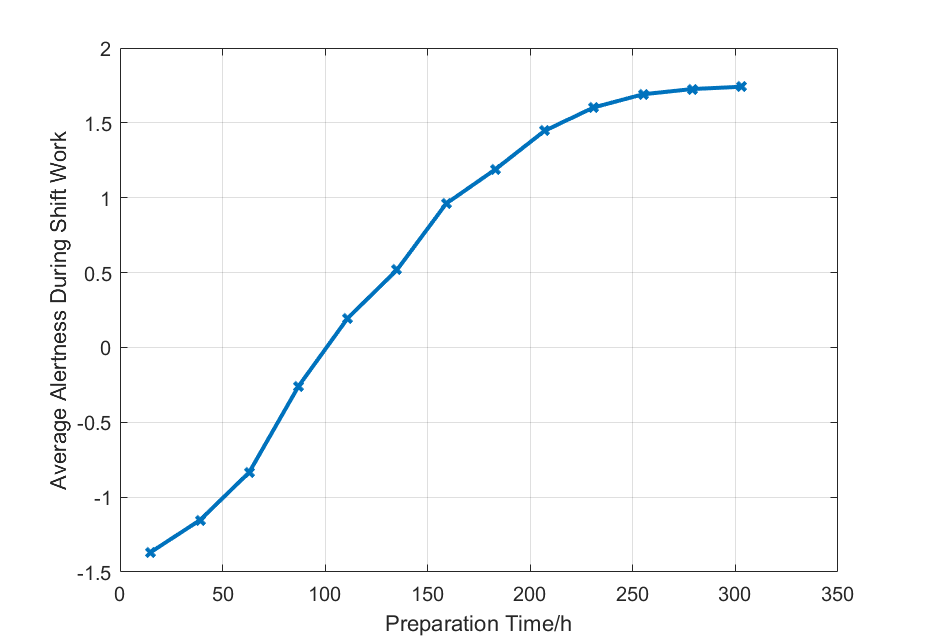}\\
(e) Average Alertness During Shift Work
\end{tabular}
\caption{Optimal schedules for 3 consecutive night shifts. The shift work time is indicated by the blue curve. The alertness $A$ is represented by the red curve. The green region indicates sleep time, and red region indicates when the light is turned on. (a) and (b) represent the light and sleep schedules which the optimization begins on the first day of shift work. (c) represents the light and sleep schedules that start 6 days ahead of the shift work. (d) represents the light and sleep schedules that start 13 days ahead of the shift work. (e) represents the number of hours the optimization begins before the first shift work versus the average alertness during the work shift. }\label{fig:Mission Opt} 
\end{figure}

Note that in \cite{yin2022human}, Yin et al. assume that sleep is only a single continuous every day. This excludes napping, a short period of sleep in addition to the usual long sleep at night. However, multiple studies have shown that taking a late nap before the night work can improve the alertness during work \cite{driskell2005efficacy,schweitzer2006laboratory}. We extend the methods in \cite{yin2022human} assuming that late nap begins when $H(t) = H^+$ and ends when $H(t) = H^-$ or work time is reached. Our simulations also demonstrate the same phenomenon for late naps. We consider three consecutive night shifts that begin at 11 PM and end at 7 AM of the next morning, and the optimization objective is to maximize alertness during three shifts. The optimization horizon begins at 8 AM on the first day and ends at 7 AM on the fourth day. One optimization uses the initial guess based on the periodic solution as we discussed and is optimized using the gradient descent algorithm. The subject will nap for 0.7 hours, 2.35 hours and 2.3 hours before the shift work, respectively. 
Another optimization uses the initial guess also based on periodic solution, but excludes the naps before the night shift. The optimal light and sleep schedules are plotted in Fig. \ref{fig:Mission Opt}. When the optimal light is $0$ lux, the subject needs to wear blocking goggles to block part of the blue spectrum \cite{rahman2011spectral}. We compare the optimized schedule of these two optimizations using average alertness $A_{avg} = \frac{J}{24}$. 
The first schedule that has naps before the shift work has an average of $A_{avg} = -1.368$ during the shift work, and the second schedule that does not have naps before the shift work has an average of $A_{avg} = -1.4533$ during the work shift. The improvement in alertness with naps before shifts can be explained by the decrease in $H$ during late napping.

However, in different scenarios, the optimal schedules will be different. If we can adjust the circadian rhythm ahead of the mission by adapting to different light and sleep schedules, alertness during shift work will be higher \cite{zhang2012optimal}. We run a series of simulations that start the optimization 1 to 13 days ahead of the first day of the shift work. The optimal schedules with 6 and 13 days of preparation are plotted in Fig. \ref{fig:Mission Opt}(c) and (d), respectively. As the preparation time increases, the average alertness during shift work also increases, as shown in Fig. \ref{fig:Mission Opt}(e). Average alertness reaches a plateau after 13 days as the circadian rhythms of the subject are fully adjusted to align with the shift work at night. Adjusting circadian rhythms provides a more significant increase in alertness, and late napping is not required in this scenario.

In summary, if the subject cannot adjust the light and sleep schedules multiple days in advance, late napping before shift work can increase average alertness during shift work. But if adjusting light and sleep schedules before shift work, the gradient descent algorithm can find better schedules that change the circadian rhythms of the subject to increase the average alertness during shift work.

\subsection{Cumulative Alertness Optimization}
In this subsection, we solve the cumulative alertness optimization problem:

Given the initial conditions $X(0) = X_0$, and the dynamics of the hybrid model represented as $\dot{X} = D_i(X,I,\beta)$ for $i = 1,\cdots, N$, light constraint Eq. \ref{eq:light con} and tunable schedule, we maximize the cumulative alertness of a person during certain period:
\begin{equation}
    \max \int_{t_0}^{t_f} (1-\beta(t)) [H^+(t)-H(t)] dt,
\end{equation}
where $t_0$ is the starting time and $t_f$ is the end time of the optimization horizon.

The gradient descent algorithm in Section \ref{shift work alertness optimization} applies to this problem with the only exception of $L$ in Eq. \ref{eq:def J} is replaced by
\begin{equation}
    L = - (1-\beta(t))[H^+(t)-H(t)].
\end{equation}

Here, we study two cases of cumulative alertness optimization. Both cases start at 12 PM and end at 12 AM on the fifth night. 

\begin{figure}[!htb]
\begin{tabular}{cc}
  \includegraphics[width=0.8\textwidth]{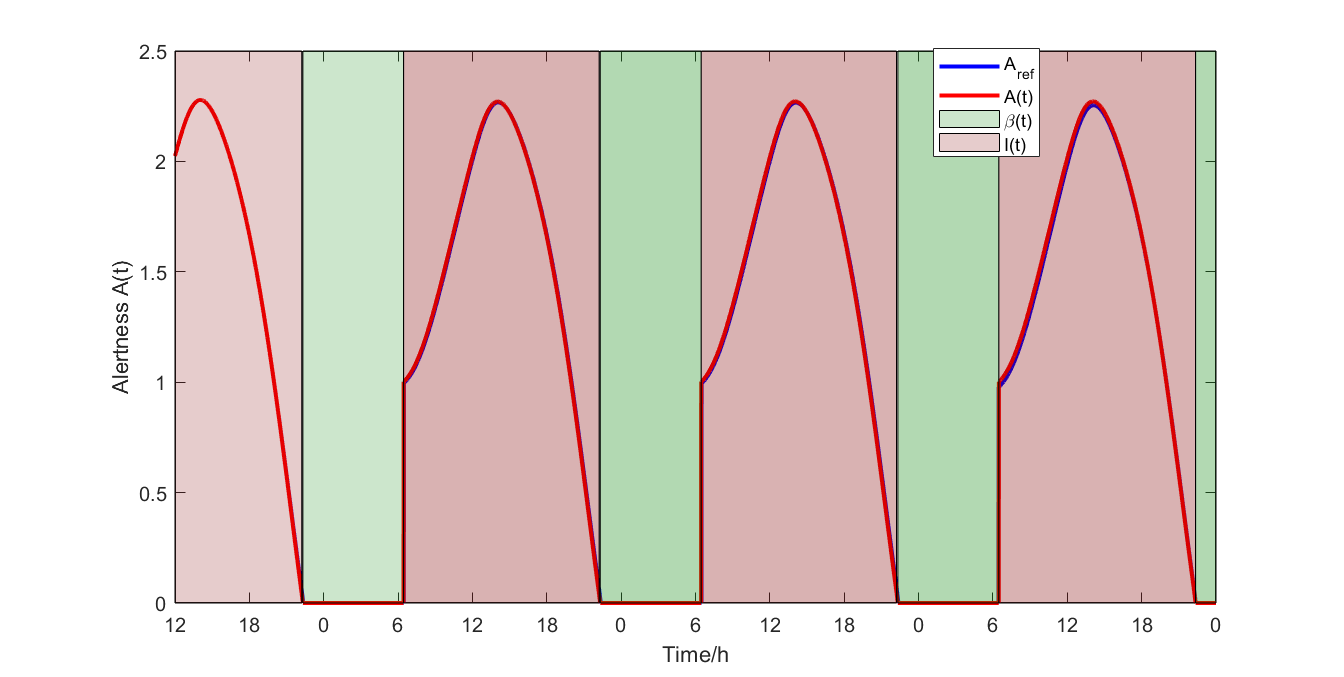}\label{fig:periodic alert opt}  \\
(a) Subject follows periodic schedule \\
\includegraphics[width=0.8\textwidth]{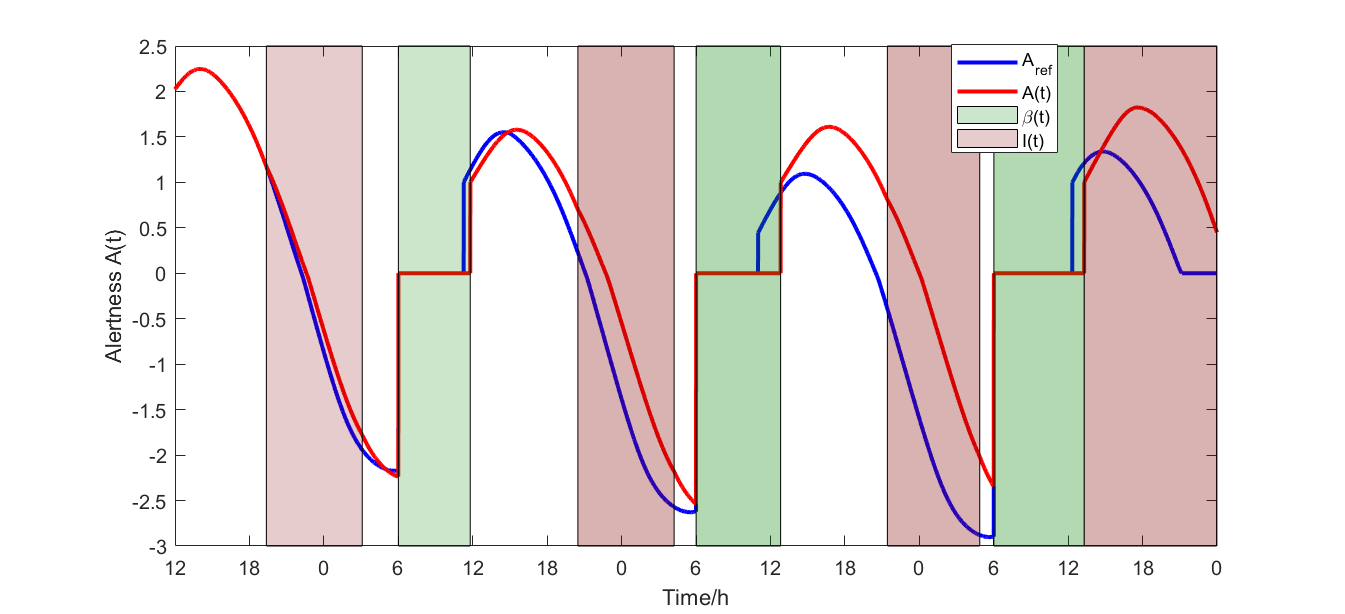}\\
(b) Subject follows 3 night shifts \\
\includegraphics[width=0.8\textwidth]{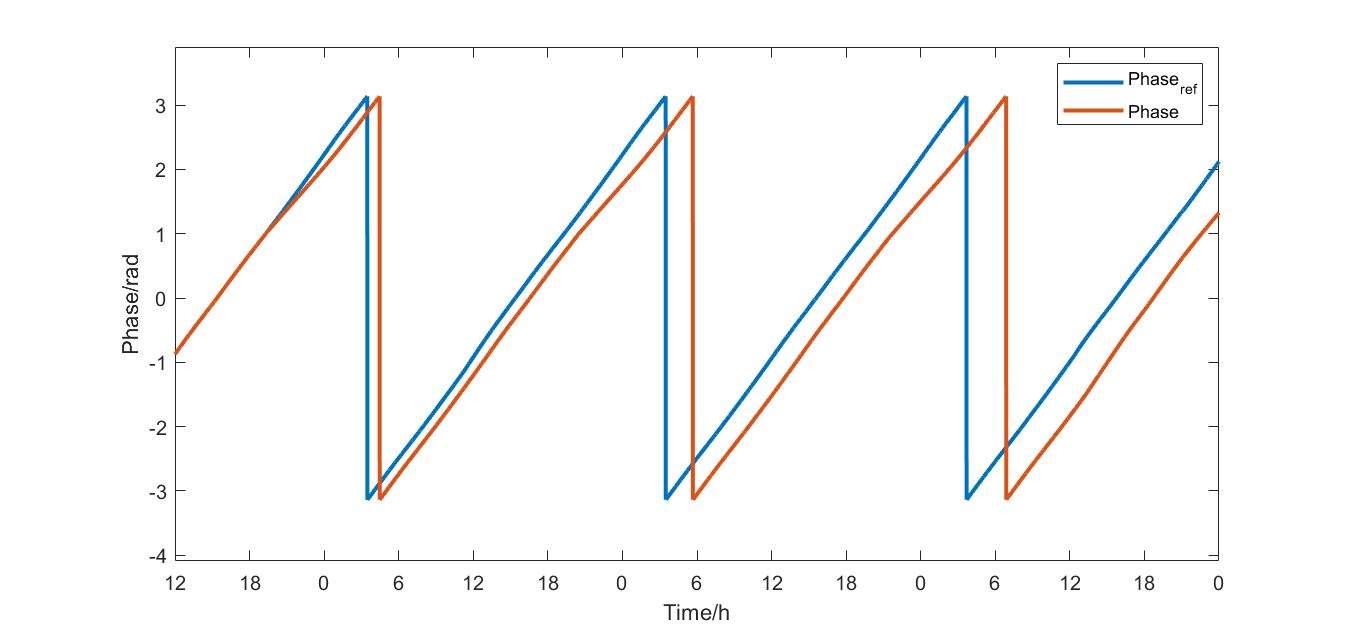}\\
(c) Phase Comparison of the Subject follows 3 night shifts \\
\end{tabular}
	\caption{Alertness comparisons of the subject before and after optimization. (a) represents the case of a subject accommodates to the periodic light and (b) represents the case that has 10 PM to 6 AM night shifts. $A_{ref}(t)$ represents the alertness of the subject before optimization. $A(t)$ represents the alertness of the subject after the optimization. The green region indicate sleep time and red region indicates when the light is turned on. (c) Circadian phase comparison of the reference schedule before optimization in red and the schedule after optimization in blue.}
	\label{fig:3 shift opt} 
\end{figure}

In the first case, we optimize the cumulative alertness of a subject that accommodates the periodic light in Eq. \ref{eq:light} and spontaneous sleep schedule. 
%The optimization starts at 12 p.m. and we optimize the cumulative alertness over the next three and a half days. 
We find that the cumulative alertness increases from 89.28 to 89.56 after optimization. The alertness comparison is plotted in Fig. \ref{fig:3 shift opt}(a). The difference is less than 0.3\%. This shows that the periodic light schedule and spontaneous sleep schedule are very close to the light and sleep schedules that maximize the cumulative alertness during the optimization horizon if there is no shift work. 
The second case has night shifts that start at 10 PM and end at 6 AM in the first three nights. After night shifts, the subject falls asleep spontaneously. We start with the initial guess based on periodic solution. The cumulative alertness is $-2.3247$ before optimization. After optimizing the light and sleep schedules, cumulative alertness increases to $27.631$. The alertness comparison is plotted in Fig. \ref{fig:3 shift opt}(b). In Fig. \ref{fig:3 shift opt}(c), the circadian phase on day 4 is delayed 3.1 hours compared to day 1. This shows that optimal light and sleep schedules delay the circadian phase to align with activities at night.

%\subsection{Relationship Between Daytime Sleepiness and Alertness}
%In this section, we compare the optimal cumulative alertness (OCA) schedule with another sleep schedule that aims to reduce daytime sleepiness and increase cognitive ability. 
In \cite{hong2021personalized}, Hong et al. use the PR model to formulate circadian necessary sleep(CNS), which is the minimum sleep duration that sleep homeostasis $H$ decreases below the sleep threshold $H^+$. Natural wake time interval (NWTI) is defined as the time such that $\{t | H(t) \leq H^+(t)\}$ so that the subject is naturally awake. If the subject's sleep time is less than CNS, then the sleep is circadian insufficient sleep (CIS) and the subject will experience more sleepiness during the wake-up time. The subject following CNS shows more NWTI than the subject following CIS.

\begin{figure}[!htb]
\begin{tabular}{cc}
  \includegraphics[width=0.8\textwidth]{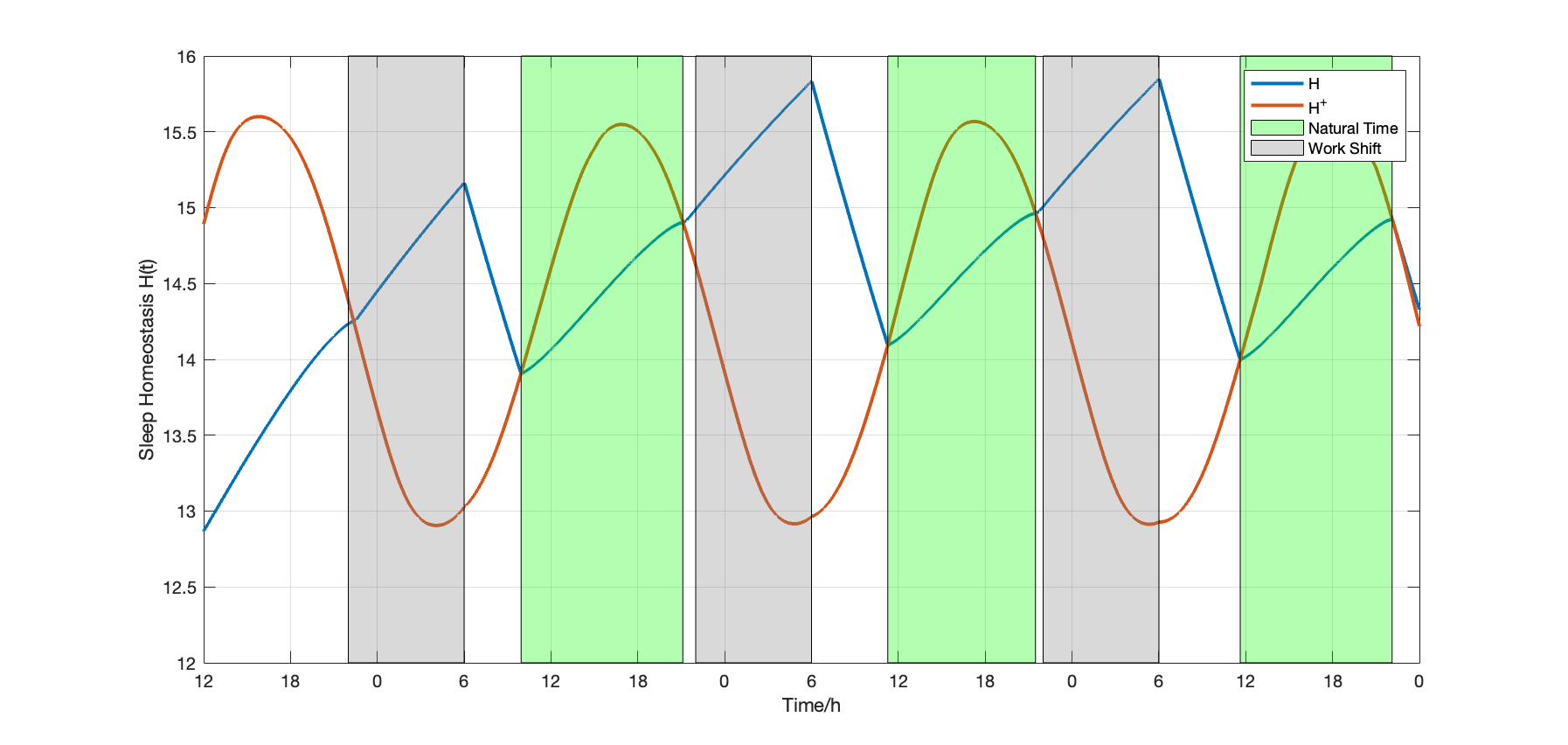}\\
(a) Subject follows CNS schedule \\
\includegraphics[width=0.8\textwidth]{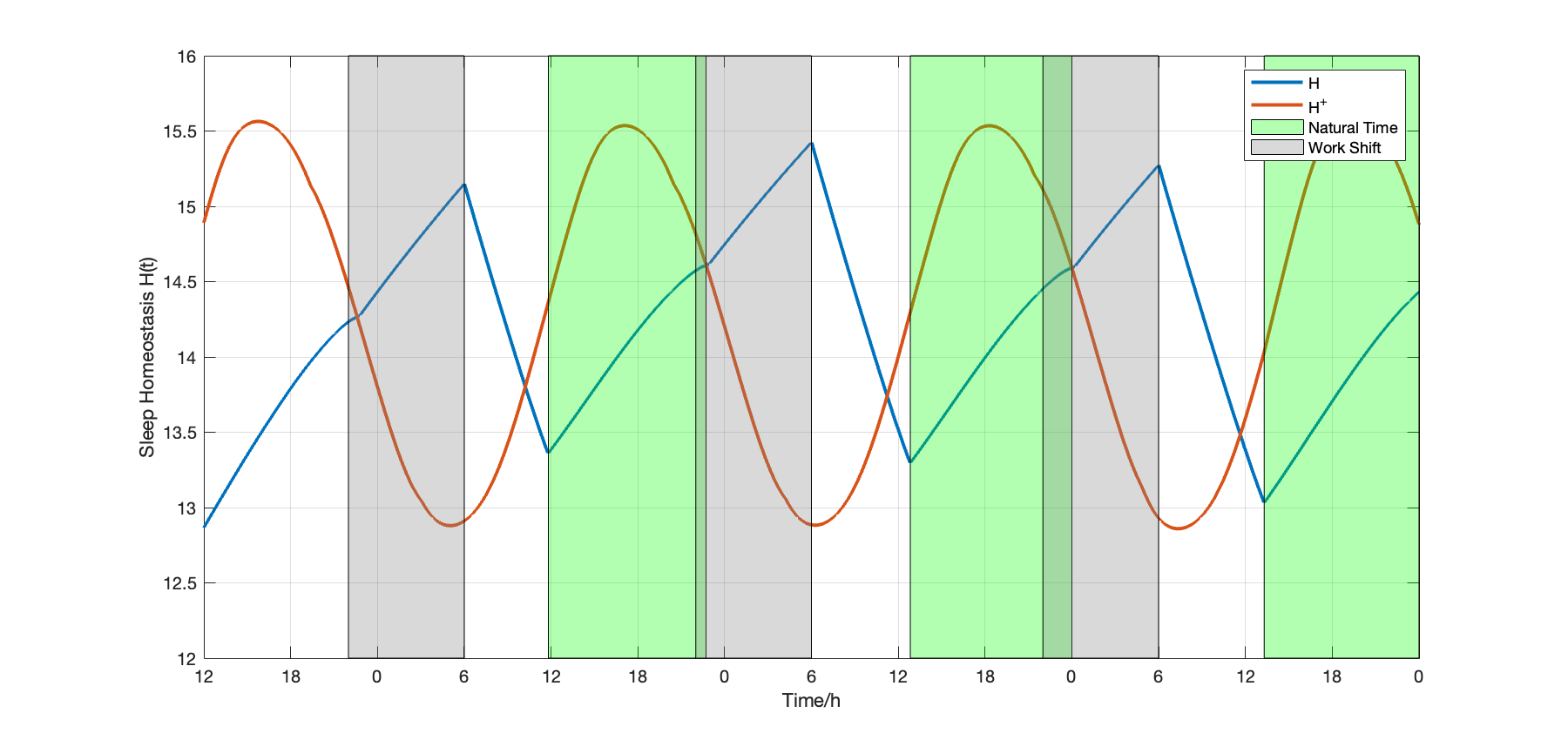}\\
(b) Subject follows optimal cumulative alertness schedule
\end{tabular}
	\caption{Natural wake time interval comparisons of the subject following CNS schedule and optimal cumulative alertness schedule. This shows the plot of alertness of the subject before and after the optimization. $A_{ref}(t)$ represents the alertness of the subject before optimization. $A(t)$ represents the alertness of the subject after the optimization. Green regions indicate the natural wake time interval and red regions indicate the shift work.}
	\label{fig:Natural time} 
\end{figure}

We hypothesize that optimizing cumulative alertness will not reduce NWTI. We run simulations with our hybrid PR model and compare NWTI with the CNS schedule. Fig. \ref{fig:Natural time} shows a comparison of NWTI between the subject following the CNS schedule and the subject following the optimal cumulative alertness schedule obtained from the gradient descent algorithm. The NWTI increases from $31.91$ hours to $32.81$ hours. We randomly sampled 3-night consecutive schedules in which every shift work begins between 3 PM and 11:00 PM and the shift duration is randomly sampled between 4 and 12 hours. The beginning and ending time of the optimization horizon is 12 PM on the first day and 12 Am on the fifth day. In 50 simulations, the average increase in cumulative alertness is 29.97. In 42 cases, the natural wake time interval was longer under the optimal cumulative alertness schedule than under the CNS schedule. The average percentage NWTI increase is 1.24\% with a standard deviation of 1.61\%. Therefore, the optimal schedules obtained from gradient descent algorithm can increase alertness during the full optimization horizon and not sacrifice NWTI.

%Song et al. propose adaptive circadian split sleep (ACSS). The ACSS pattern includes the main sleep period that follows the CNS and a brief nap taken before the start of shift work.  

\section{Conclusion}
This paper proposes the hybrid PR model that consists of the physiology underlying sleep-wake dynamics and can accurately predict cognitive abilities better than the three-process model. The hybrid model does not have stiff equations in the full PR model and has continuous gradients. We apply the calculus of variation to find the light and sleep schedule that maximizes cumulative alertness during shift work and the full optimization horizon. We find that if we can adjust the circadian rhythms before shift work, alertness during shift will be higher than not adjusting the circadian rhythm. If we cannot adjust the circadian rhythms beforehand, taking a late nap before the shift will increase alertness during shift work. We also compare the optimal cumulative alertness schedule with the CNS schedule and demonstrate that maximizing cumulative alertness during the full optimization horizon does not sacrifice the natural wake time interval.

\section{Appendix}
\label{appendix}
\subsection{Calculus of Variation Derivation}
Suppose the dynamics for the $i$-th mode is described by the following dynamics:
\begin{equation}
    \dot{X} = D_i(X,I), t\in[t_{i-1},t_i),\forall i \in\{1,2,\cdots,N\},
\end{equation}
where $N$ is the total number of modes during the optimization time horizon. $t_0 = 0$ and $t_N = t_f$ are fixed. All the switching time $t_i$ are subject to the inequality constraints in tunable schedule in Eq. \ref{eq:tunable constr}. An objective function is formulated as 
\begin{equation}
    J = \int_0^{t_f} L(t,X,I)dt.
\end{equation}
We introduce Lagrange multipliers $\lambda(t)\in \mathbf{R}^4$.
The augmented objective function is:
\begin{align}
    J_a = \int_0^{t_f} L(\tau,X,I) d\tau + \sum_{i=1}^{N} \int_{t_{i-1}}^{t_i} \lambda(\tau)[D_i(X,I)-\dot{X}(\tau)] d\tau.\label{eq:aug}
\end{align}
We introduce a small scalar $\alpha$ and add perturbations to the inputs $\alpha \delta I$ and $\alpha \delta t_i$. The perturbed augmented objective function becomes:
\begin{equation}
    \begin{split}
    J_a + \alpha \delta J_a + o(\alpha) = \int_0^{t_f} L(\tau,X+\alpha \delta X + o(\alpha),I+\alpha \delta I) d\tau \\
    + \sum_{i=1}^{N}\int_{t_{i-1}+\alpha \delta t_{i-1}}^{t_i+\alpha \delta t_{i}} \lambda^T(\tau) [D_i(X+\alpha \delta X + o(\alpha),I+\alpha \delta I-\dot{X} - \alpha \delta \dot{X} - o(\alpha)] d\tau,
    \end{split}\label{eq:per aug}
\end{equation}
where  $\delta X, \delta J_a$ are the first order perturbation in states and augmented cost, and $o(\alpha)$ represents the terms that has higher order than first order of $\alpha$.
Formally,
\begin{equation}
    \delta X = \lim_{\alpha \rightarrow 0} \frac{X(I(t)+\alpha \delta I(t),t_i+\alpha \delta t_i)-X(I(t),t_i)}{\alpha},
    \label{eq:X var}
\end{equation}
\begin{equation}
    \delta J_a = \lim_{\alpha \rightarrow 0} \frac{J_a(I(t)+\alpha \delta I(t),t_i+\alpha \delta t_i)-J_a(I(t),t_i)}{\alpha}.
    \label{eq:Ja var}
\end{equation}
The first variation of the augmented cost is calculated by subtracting the perturbed objective function Eq. \ref{eq:per aug} by Eq. \ref{eq:aug} and taking the limit $\alpha \rightarrow 0$.
\begin{equation}
    \begin{split}
        \delta J_a &= \int_{t_0}^{t_N} \left\{\left[\frac{\delta L(\tau,X,I)}{\delta X}\right]^T \delta X + \frac{\delta L(\tau,X,I)}{\delta I} \delta I \right\} d \tau\\
        &+ \sum_{i=0}^{N} \int_{t_{i-1}}^{t_i} \left\{\dot{\lambda}^T \delta X + \lambda^T (\tau) \left[\left( \frac{\delta D_i(X,i)}{\delta X}\right)^T \delta X + \frac{\delta D_i(X,I)}{\delta I} \delta I \right] \right\} d \tau\\
        & - \sum_{i = 1}^{N} \left[\lambda^T (t_i^-) \delta X(t_i^-) - \lambda^T (t_{i-1}^+) \delta X(t_{i-1}^+)\right]\\
        & + \sum_{i = 1}^{N-1} \left[\lambda^T(t_i^-)D_i(X(t_i^-,I(t_i^-)))-\lambda^T(t_i^+)F_{i+1}(X(t_i^+),I(t_i^+))\right] \delta \tau_i\\
        &+ \sum_{i = 1}^{N-1} \left[L(t_i^-,X(t_i^-),I(t_i^-)) - L(t_i^+,X(t_i^+),I(t_i^+))\right] \delta \tau_i
    \end{split}
\end{equation}
By selecting the Lagrange multiplier $\lambda(t)$ as:
\begin{align}
    &\lambda(t_N) = 0,&\\
    \dot{\lambda}(t)=-\frac{\delta L(t,X,I)}{\delta X} - &\left(\frac{\delta D_i(X,I)}{\delta X}\right)^T \lambda(t) \textrm{, when} t\in [t_{i-1},t_i),\\
    &\lambda(t_i^-) = \lambda(t_i^+),
\end{align}
$\delta J_a$ can be simplified to
\begin{equation}
    \begin{split}
        \delta J_a = \sum_{i=1}^N \int_{t_{i-1}}^{t_i} \left[\frac{\delta L(\tau,X,I)}{\delta I} + \lambda^T(\tau)\frac{\delta D_i(X,I)}{\delta I} \right]^T  \delta I d \tau\\
        + \sum_{i=1}^{N-1}[L(t_i^-,X(t_i^-),I(t_i^-))-L(t_i^+,X(t_i^+),I(t_i^+))+ \lambda(t_i^-) D_i(X(t_i^-),I(t_i^-))\\- \lambda^T(t_i^+) D_{i+1}(X(t_i^+),I(t_i^+))]\delta t_i.
    \end{split}\label{eq:simplified first var}
\end{equation}
From Eq. \ref{eq:simplified first var}, the gradient of $J$ with respect to $I$ and $t_i$ are given as 
\begin{equation}
    \nabla_{I(t)} J = \frac{\partial L(t,X,I)}{\partial I} + \lambda^T(t) \frac{\partial D_i(X,I)}{\partial I},
\end{equation}
\begin{equation}
\begin{split}
    \nabla_{t_i}J = &L(t_i^-,X(t_i^-),I(t_i^-))-L(t_i^+,X(t_i^+),I(t_i^+))+ \lambda(t_i^-) D_i(X(t_i^-),I(t_i^-)) %\label{eq:grad I}
    \\&- \lambda^T(t_i^+) D_{i+1}(X(t_i^+),I(t_i^+))
\end{split}
\end{equation}

\bibliographystyle{plos2015}
\bibliography{ref}

\end{document}